\documentclass[a4paper,11pt]{article}
\usepackage{jcappub}
\usepackage{amsthm}
\usepackage{graphicx}
\usepackage{color}
\usepackage{graphicx,textcomp,float,slashed}
\usepackage{hyperref}
\usepackage{epstopdf}
\usepackage{multirow}

\newcommand{\qVis}{q_\text{vis}}
\newcommand{\QVis}{Q_\text{vis}}
\newcommand{\tBBN}{t_\text{BBN}}
\newcommand{\TBBN}{T_\text{BBN}}
\newcommand{\newc}{\newcommand}
\newc{\gsim}{\lower.7ex\hbox{$\;\stackrel{\textstyle>}{\sim}\;$}}
\newc{\lsim}{\lower.7ex\hbox{$\;\stackrel{\textstyle<}{\sim}\;$}}
\newc{\gev}{\,{\rm GeV}}
\newc{\mev}{\,{\rm MeV}}
\newc{\ev}{\,{\rm eV}}
\newc{\kev}{\,{\rm keV}}
\newc{\tev}{\,{\rm TeV}}

\newc{\mz}{M_Z}
\newc{\mw}{m_{\rm weak}}
\newc{\nr}[1]{N^c_R{}_{#1}}

\usepackage{amsmath}
\def\beq{\begin{equation}}
\def\eeq{\end{equation}}
\def\bea{\begin{eqnarray}}
\def\eea{\end{eqnarray}}
\def\bitem{\begin{itemize}}
\def\eitem{\end{itemize}}
\newcommand{\bec}{\begin{center}}
\newcommand{\eec}{\end{center}}
\def\bar#1{\overline{#1}}

\def\abs#1{\left| #1\right|}

\def\inv{^{\raise.15ex\hbox{${\scriptscriptstyle -}$}\kern-.05em 1}}
\def\lbar{{\lower.35ex\hbox{$\mathchar'26$}\mkern-10mu\lambda}} %lambda bar

\def\om#1#2{\omega^{#1}{}_{#2}}

\def\to{\rightarrow}

\let\<=\langle
\let\>=\rangle

\let\+=\uparrow

\def\be{\beta}
\def\ga{\gamma}

\def\la{\lambda}

\def\si{\sigma}

\def\om{\omega}
\def\De{\Delta}
\def\Ga{\Gamma}

\def\La{\Lambda}

\def\Om{\Omega}

\newcommand{\ben}{\begin{equation}}
\newcommand{\een}{\end{equation}}
\newcommand{\ba}{\begin{array}}
\newcommand{\ea}{\end{array}}
\newcommand{\bit}{\begin{itemize}}
\newcommand{\eit}{\end{itemize}}

\newcommand{\dm}{dark matter }			% abbreviation or spell out?
\newcommand{\Mpl}{M_{\rm Pl}}			% Planck mass
\newcommand{\mpl}{m_{\rm Pl}}			% Reduced Planck mass
\newcommand{\mH}{m_{H^0}}				% Mass of H0 particles
\newcommand{\Ebar}{\bar{E}}			% Average energy of H0 particles from defect decay
\newcommand{\refsub}{\chi}				% Reference subscript
\newcommand{\tref}{t_\refsub}	                            % Reference time
\def\beq{\begin{equation}}
\def\eeq{\end{equation}}

\let\be=\beta
\let\ga=\gamma
\let\Ga=\Gamma

\let\De=\Delta

\let\la=\lambda
\let\La=\Lambda

\let\si=\sigma

\let\om=\omega
\let\Om=\Omega

\numberwithin{equation}{section}

\usepackage{amsmath,amsfonts,amssymb}
\usepackage[a4paper]{geometry}
\usepackage{multicol}
\usepackage{graphicx}
\usepackage{float}

\title{ \Huge Dark Matter with Topological Defects in the Inert Doublet Model}

\author[1,2]{Mark~Hindmarsh,}
\author[3]{Russell~Kirk,}
\author[1]{Jose Miguel No,}
\author[3]{and Stephen M. West}

\affiliation[1]{Dept. of Physics and Astronomy, University of Sussex, 
Brighton BN1 9QH, U.K.}
\affiliation[2]{Department of Physics and Helsinki Institute of Physics, P.O.\ Box 64, 00014 Helsinki University, Finland} 
\affiliation[3]{Dept. of Physics,
Royal Holloway University of London, Egham, Surrey TW20 0EX, U.K.}
\emailAdd{m.b.hindmarsh@sussex.ac.uk}
\emailAdd{russell.kirk.2008@live.rhul.ac.uk}
\emailAdd{J.M.No@sussex.ac.uk}
\emailAdd{stephen.west@rhul.ac.uk}

\abstract{We examine the production of \dm by decaying topological defects in the high mass region $m_{\mathrm{DM}} \gg m_W$ of the Inert Doublet Model, extended with an extra U(1) gauge symmetry. The density of \dm states (the neutral Higgs states of the inert doublet) is determined by the interplay of the freeze-out mechanism and the additional production of \dm states from the decays of topological defects, in this case cosmic strings. These decays increase the predicted relic abundance compared to the standard freeze-out only case, and as a consequence the viable parameter space of the Inert Doublet Model can be widened substantially. In particular, for a given \dm annihilation rate lower \dm masses become viable.
We investigate the allowed mass range taking into account constraints on the energy injection rate from the diffuse $\ga$-ray background and Big Bang Nucleosynthesis, together with constraints on the \dm properties coming from direct and indirect detection limits. For the Inert Doublet Model high-mass region, an inert Higgs mass as low as $\sim 200$ GeV is permitted. 
There is also an upper limit on string mass per unit length, and hence the symmetry breaking scale, from the relic abundance in this scenario. Depending on assumptions made about the string decays, the limits are in the range $10^{12}$ GeV to $10^{13}$ GeV.
}

\begin{document}

\maketitle

\vspace{1cm}
\section{Introduction}
%%%%%%%%%%%%%%%%%%%%%%%%%%%%%%%%%%%%%%%%%%%%%%%%%%%%%%%%

The identity of \dm remains one of the major puzzles in modern physics.  With a relic abundance of $\Om_{\rm DM}h^2 = 0.1186\pm 0.0031$ as measured by Planck \cite{Ade:2013ktc}, \dm constitutes around five times more of the energy density of the Universe than normal baryonic matter. It is also one of the clearest signs that there is physics beyond the Standard Model (SM). Nothing in the SM can play the role of \dm and fully explain the measured value for the relic abundance and so we are forced to look beyond it for candidates. 

In extending the SM from a bottom up perspective, we need to add dark matter states, and we may also wish to supplement the SM gauge group with additional gauge symmetries.  The simplest example is the addition of an Abelian gauge symmetry, which we will refer to as $U(1)^{\prime}$. The associated gauge boson can play a role in connecting the \dm state to the SM sector \cite{ArkaniHamed:2008qp}. Such a connection is required in most models of \dm relying on the freeze-out mechanism to determine the relic abundance. 

We may also expect additional Abelian gauge symmetries from a top down perspective. Whether one considers heterotic strings, type II string theory with D-branes or F-theory, it is common that in attempts to recover the SM gauge group, additional unbroken Abelian gauge symmetries are generated (see for example \cite{Abel:2008ai,Goodsell:2010ie,Dudas:2010zb}). 

Whatever the source of the extra symmetry, the $Z^{\prime}$ in these models must be sufficiently massive to have escaped detection, and so the $U(1)^{\prime}$ must be spontaneously broken. During the resulting phase transition in the early Universe, linear topological defects (TDs) or cosmic strings would have been formed \cite{Kibble:1976sj}. The mass per unit length $\mu$ of the strings is proportional to the square of the symmetry breaking scale $v^{\prime}$.

The strings decay into either the particles of their constituent fields or gravitational radiation (see {\it e.g.} \cite{Hindmarsh:2011qj}). The specific branching fractions into each are uncertain and are usually left as an unknown parameter of the model. A number of existing works have examined the constraints on these cosmic strings for a range of scenarios, symmetry breaking scales and constituent fields \cite{Jeannerot:1999yn,Hindmarsh:2013jha,Hyde:2013fia,Mota:2014uka,Long:2014mxa}. 

The connection between \dm and the TDs can also be exploited to find constraints on the properties of the cosmic strings.  Whether the \dm states are charged under the additional symmetry or not, we generically expect that the decays of the TDs formed in these models will have a branching fraction to the \dm states at some level. An intriguing alternate connection between axion-like domain walls and dark matter has been discussed in \cite{Derevianko:2013oaa,Pospelov:2012mt,Stadnik:2014cea}.

There are number of ways in which the connection can be realised. If the \dm is not charged under the $U(1)'$, but is a scalar state, $\chi$, it can couple to the complex scalar, $\phi$, responsible for spontaneously breaking the $U(1)^{\prime}$. It can do so via a quartic ``portal" coupling,  $\mathcal{L}\supset \abs{\phi}^2\abs{\chi}^2$. This portal coupling will provide a connection between the \dm states and the states that will form the cosmic string.

A second example is where the additional $U(1)^{\prime}$ kinetically mixes with the SM gauge group $U(1)_Y$ (which could allow for a further channel for particle radiation from the string \cite{Vachaspati:2009jx}). If the \dm is charged under the electroweak gauge group we again have a direct connection between the $U(1)'$ sector and the \dm states. 

A further example, is where the \dm state is charged under the $U(1)'$. In this case the connection is straightforward with the \dm states being produced directly from the decays of the TDs. In all these cases it is clear that \dm will be produced in the decays of these TDs with some branching fraction.

In \cite{Hindmarsh:2013jha} this scenario was examined in a model independent way with the \dm injection rate from the defect decays varying as a power law with time.\footnote{Dark matter from decaying strings was also considered in \cite{Jeannerot:1999yn} where a specific production mechanism was assumed, in which \dm particles were generated, in small numbers, only when the loops of string had shrunk to radii the same order as the string width.}

It is clear that if we increase the injection of \dm states from TD decays, the annihilation cross section needs to increase in order to bring the relic abundance down to the measured value. We can only increase the annihilation cross section up to the unitarity limit and hence there is an upper bound on the \dm injection rate.
For a given \dm mass, this translates into constraints on the properties of the cosmic string network, and specifically on the mass per unit length of the strings.
A further effect of increasing the annihilation rate is generically an enhancement of the indirect and direct detection signals, which can further limit the model. 

Conversely, the extra source of \dm particles changes the predictions of specific \dm models. 

Scenarios in which \dm production by decaying TDs could play a positive role involve weakly interacting massive particle (WIMP) models where \dm annihilation into SM particles naturally yields a relic density significantly below the observed one. This is, for example, the case for WIMP scenarios where \dm annihilates dominantly via $SU(2)_L\times U(1)_Y$ gauge interactions, as in most of the parameter space of the minimal \dm \cite{Cirelli:2005uq,Cirelli:2007xd} and IDM \cite{Ma:2006km,Barbieri:2006dq,LopezHonorez:2006gr} scenarios.
These models constitute rather minimal extensions of the SM, and are very appealing phenomenologically since the \dm annihilation properties can be purely dictated by its $SU(2)_L\times U(1)_Y$ gauge quantum numbers. However, this same feature greatly restricts the range of \dm masses that yield the observed relic abundance via thermal freeze-out. 

In this work we explore the impact of \dm production by the decay of TDs on the available parameter space for these scenarios, taking as a case study the Inert Doublet Model (IDM) supplemented by an additional $U(1)^{\prime}$ gauge symmetry. This model has already been considered in \cite{Ko:2014uka,Ko:2014eqa}, with the different motivation of including  in a gauge group the discrete symmetry stabilising the \dm particle. The spontaneous breaking of this $U(1)^{\prime}$ gauge symmetry in the early universe leads to the formation of cosmic strings whose decay represents the extra source of \dm production. We will show that this greatly opens-up the allowed range of \dm masses in the IDM. 

Before exploring this in detail, in Section \ref{IDMReview} we review the main features of the IDM that will be relevant for the subsequent discussion. In Section \ref{sec:u1ext} we extend the IDM by a $U(1)'$ gauge symmetry and analyse its implications. In Section \ref{CSSection} we discuss and parameterise the injection rate of \dm via the decay of cosmic strings. Following this, in Section \ref{sec:boltzsolve} we analyse the impact of the \dm production by cosmic strings on the parameter space of the IDM by solving the relevant Boltzmann equations, and discuss the connection between the mass of the \dm and the scale of $U(1)^{\prime}$ symmetry breaking. In Section \ref{sec:DefectConstraints}, we discuss observational constraints on the presence of these TDs and finally, we conclude in Section \ref{sec:cons}.

\section{The Inert Doublet Model: A brief review}\label{IDMReview}
%%%%%%%%%%%%%%%%%%%%%%%%%%%%%%%%%%%%%%%%%%%%%%%%%%%%%%%%

The SM contains a single Higgs doublet $H_1$ whose vacuum expectation value (VEV) breaks the $SU(2)_L \times U(1)_Y$ symmetry down to $U(1)_{\rm{EM}}$. The IDM \cite{Ma:2006km,Barbieri:2006dq,LopezHonorez:2006gr} extends the SM by adding a second Higgs doublet $H_2$ which is odd under an imposed $\mathbb{Z}_2$ symmetry, with all SM states even under this symmetry. The extra Higgs doublet does not couple to fermions and does not acquire a VEV ($\mu_2^2 > 0$). The lightest neutral component of this {\it inert} doublet, is stable and can therefore be a potentially viable \dm candidate. The most general scalar potential one can then write for $H_1$, $H_2$ is
\beq\label{Potential}
\begin{split}
V = & -\mu_1^2 |H_1|^2 + \mu_2^2 |H_2|^2 + \lambda_1 |H_1|^4 + \lambda_2 |H_2|^4 + \lambda_3 |H_1|^2|H_2|^2 \\& + \lambda_4 |H_1^{\dagger}H_2|^2 + \frac{\lambda_5}{2}\left\{(H_1^{\dagger} H_2)^2 + h.c.\right\},
\end{split}
\eeq
with
\begin{equation}\label{FieldDefs}
H_1 = \begin{pmatrix}0\\ \frac{1}{\sqrt{2}}(v + h)\end{pmatrix}, \,\,\,\,\,\,\, H_2 = \begin{pmatrix}H^+\\ \frac{1}{\sqrt{2}}(H^0 + i A^0)\end{pmatrix},
\end{equation}
in the unitary gauge, where the SM Higgs VEV $v = 246$ GeV. The scalar sector is then comprised of the SM Higgs boson $h$ and four new $\mathbb{Z}_2$-odd particles ($H^{\pm}$, $H^0$ and $A^0$), with masses given by
\beq\label{MH0Mass}
m_{H^0}^2 = \mu_2^2 + \frac{1}{2}\left(\lambda_3 + \lambda_4 + \lambda_5 \right)v^2,
\eeq
\beq\label{MA0Mass}
m_{A^0}^2 = \mu_2^2 + \frac{1}{2}\left(\lambda_3 + \lambda_4 - \lambda_5 \right)v^2,
\eeq
\beq\label{MHCMass}
m_{H^{\pm}}^2 = \mu_2^2 +\frac{1}{2}\, \lambda_3\, v^2.
\eeq
We take $H^0$ to be the lightest and thus our \dm candidate, which amounts to requiring $\lambda_4+\lambda_5 < 0$ and $\lambda_5 < 0$.
We also define $\lambda_3 + \lambda_4 + \lambda_5  \equiv 2\,\lambda_L$ for later convenience. The squared mass differences among the new states are given by
\beq
\label{massD0}
\De m_0^2 \equiv m_{A^0}^2 - m_{H^0}^2 = -\lambda_5 \,v^2 > 0,
\eeq
\beq
\label{massD+}
\De m_+^2 \equiv m_{H^\pm}^2 - m_{H^0}^2 = -\frac{1}{2}(\lambda_4 + \lambda_5)\,v^2 >0.
\eeq

From theoretical constraints we can already restrict some of the parameters of the model. Firstly, the potential (\ref{Potential}) is bounded from below if
\beq\label{PotentialBounds}
\lambda_{1,2} > 0, \quad \lambda_3 > -2\sqrt{\lambda_1\,\lambda_2}, \quad \lambda_3 + \lambda_4 -\left|\lambda_5 \right| = 2\,\lambda_L  > -2\sqrt{\lambda_1\,\lambda_2}.
\eeq
We also have a condition for the global EW minimum to preserve the $\mathbb{Z}_2$ symmetry given by
\beq\label{PotentialBounds2}
\frac{\mu_1^2}{\sqrt{\lambda_{1}}} < \frac{\mu_2^2}{\sqrt{\lambda_{2}}}.
\eeq
It is also possible to derive unitarity constraints on tree-level processes among the various scalars, which read \cite{Arhrib:2012ia} (see also
\cite{Kanemura:1993hm,Ginzburg:2005dt})
\beq\label{Unitarity1}
\left|e_i \right| \leq 8 \pi \quad \forall i=1,...,12
\eeq
with
\beq\label{Unitarity2}
e_{1,2} = \lambda_3 \pm \lambda_4, \quad \quad e_{3,4} = \lambda_3 \pm \lambda_5, \quad \quad e_{5,6} = \lambda_3 +2 \lambda_4 \pm 3 \lambda_5,
\eeq
\beq\label{Unitarity3}
e_{7,8} = -\lambda_1 - \lambda_2 \pm \sqrt{ \left(\lambda_1-\lambda_2\right)^2 + \lambda^2_4}, \quad \quad e_{9,10} = -\lambda_1 - \lambda_2 \pm \sqrt{ \left(\lambda_1-\lambda_2\right)^2 + \lambda^2_5},
\eeq
\beq\label{Unitarity4}
e_{11,12} = -3\lambda_1 - 3\lambda_2 \pm \sqrt{9\left(\lambda_1-\lambda_2\right)^2 +\left(2\lambda_3+\lambda_4\right)^2}.
\eeq

\subsection{Electroweak precision observables: $S$, $T$, $U$}
\label{EWPOsec}

The new states $H^{\pm}$, $A^0$ and $H^0$ contribute to electroweak precision observables (EWPO) via loop corrections to the oblique parameters $S$, $T$, and $U$ \cite{Peskin:1990zt,Peskin:1991sw}.  
Under the assumption $U = 0$, the best-fit values and standard deviations for $S$ and $T$ from the up-to-date global analysis of EW precision observables performed by the GFitter Group \cite{Baak:2014ora}, for a SM reference point with $m_t = 173$ GeV and a $125$ GeV Higgs mass, are 
\beq
\label{chi_EWPO1}
\Delta S \equiv S - S_{\mathrm{SM}} = 0.06 \pm 0.09, \quad \quad \quad \quad
\Delta T \equiv T - T_{\mathrm{SM}} = 0.10 \pm 0.07.
\eeq
The IDM contribution to the $S$-parameter is given by
\beq\label{Sparameter}
\Delta S = \frac{1}{2\pi}\int_0^1 dx \, x\,(1-x) \, \mathrm{log} \left(\frac{x\,m^2_{H^{0}} + (1-x) m^2_{A^{0}}}{m^2_{H^{\pm}}} \right).
\eeq 
When $m_{H^\pm} \simeq m_{A^0}$ (as preferred by the $T$-parameter, see below), we have 
\beq\label{Sparameter2}
\Delta S \simeq \frac{1}{2\pi}\int_0^1 dx \, x\,(1-x) \, \mathrm{log} \left(1 - x\,a^2\right) \leq 0, \quad \quad a^2 \equiv \frac{m^2_{A^{0}}-m^2_{H^{0}}}{m^2_{A^{0}}} \in [0,1],
\eeq 
where $\Delta S$ is monotonically decreasing with $a^2$ and approaches $\Delta S \simeq - 0.022$ for $a^2 \to 1$, well within the range favoured by the global fit (\ref{chi_EWPO1}).
Thus, we can safely disregard the $S$-parameter in the following discussion. The most important IDM contribution affects the $T$-parameter \cite{Barbieri:2006dq}:
\beq\label{Tparameter}
\Delta T = \frac{\left[ F\left(m_{H^{\pm}},m_{A^{0}} \right) + F\left(m_{H^{\pm}},m_{H^{0}} \right) - F\left(m_{A^{0}},m_{H^{0}} \right)\right]}{32\pi^2\alpha_{\mathrm{EM}} v^2},
\eeq
with
\beq\label{Tparameter2}
F\left(m_1,m_2 \right) = \frac{m_1^2+m_2^2}{2} - \frac{m_1^2\, m_2^2}{m_1^2 - m_2^2}\, \mathrm{log} \left(\frac{m_1^2}{m_2^2} \right) \, .
\eeq 
Noting that $F\left(m_2,m_1 \right) = F\left(m_1,m_2 \right)$ and $F\left(m_1,m_1 \right) = 0$, we immediately obtain from (\ref{Tparameter}) that $\Delta T = 0$ in the IDM for either $m_{A^0} = m_{H^\pm}$ or  $m_{H^0} = m_{H^\pm}$. This can be understood  (see {\it e.g.} \cite{Kanemura:1996eq} and Appendix \ref{AppA} for details) by recasting the potential (\ref{Potential}) in terms of the $2\times2$ matrices ${\bf\Phi}_{1}=(i\sigma _{2}H_{1}^{*}, H_{1}), \, {\bf\Phi}_{2}=(i\sigma _{2}H_{2}^{*}, \pm H_{2})$, which then preserves a global $SU(2)_{L}\times SU(2)_{R}$ symmetry in the limit $\lambda_4 = \pm\lambda_5$. By virtue of (\ref{massD0}) and (\ref{massD+}), this {\it custodial} symmetry translates into either $\Delta m^2_{+} = \Delta m^2_{0}$ or $\Delta m^2_{+} = 0$.

\section{An Inert Doublet Model with a $U(1)'$ gauge symmetry}\label{sec:u1ext}
%%%%%%%%%%%%%%%%%%%%%%%%%%%%%%%%%%%%%%%%%%%%%%%%%%%%%%%%

As we have discussed above, we may expect there to be additional Abelian gauge symmetries that are broken at some scale $v'$ whether this is desired from a bottom up perspective or whether these are remnants from a UV theory. In context of the IDM, we can also motivate a $U(1)^{\prime}$. We require a stabilising symmetry, $\mathbb{Z}_2$, for the stability of our \dm state and this can be generated as a remnant of this additional $U(1)^{\prime}$ after it is spontaneously broken. In any case, 
in order to spontaneously break this symmetry we introduce a new complex scalar, $\phi$, which has some charge under the $U(1)^{\prime}$.

The phase transition associated with this breaking in the early universe gives rise to cosmic strings, which can then couple to the inert doublet $H_2$.
This coupling is generated from the following scalar potential
\beq
\label{Potential2}
\begin{split}
V = & -\mu_1^2 |H_1|^2 + \mu_2^2 |H_2|^2 - \mu_{\phi}^2 |\phi|^2 + \lambda_1 |H_1|^4+ \lambda_2 |H_2|^4 +\lambda_{\phi} |\phi|^4 + \tilde{\lambda}_1 |\phi|^2|H_1|^2  \\& 
+ \tilde{\lambda}_2 |\phi|^2|H_2|^2 + \lambda_3 |H_1|^2|H_2|^2 + \lambda_4 |H_1^{\dagger}H_2|^2 + \left\{\frac{\lambda_5}{2}(H_1^{\dagger} H_2)^2 + h.c.\right\}.
\end{split}
\eeq
The field $\phi$ gains an expectation value spontaneously breaking the $U(1)^{\prime}$ and as usual we expand $\phi$ about its VEV in the unitary gauge as
\beq
\phi = \frac{1}{\sqrt{2}}\left(v'+X\right) \,.
\eeq
We have a choice about whether the inert doublet is charged under the $U(1)^{\prime}$ or not. This has a consequence for the generation of the $\la_5$ term in Equation~\ref{Potential2}. If we charge $H_2$ (and not $H_1$), then the $\la_5$ term is forbidden. In the absence of this term, $H^0$ and $A^0$ are degenerate (see (\ref{massD0})). This allows $H^0 N \to A^0 N$ scattering (N being a nucleon) via $Z$-boson exchange to occur and this leads to a direct dark matter detection signal much larger than current experimental bounds, which rules out $\lambda_5 = 0$ \cite{Barbieri:2006dq}. 

To avoid this issue, a $\lambda_5$ term could be generated through a higher-dimensional effective operator, e.g.
\beq
\frac{1}{\La}\phi\,(H_1^{\dagger} H_2)^2,
\eeq
where $\La$ parameterises some high scale physics that has been integrated out. This operator is allowed if $H_2$ and $\phi$ have charges $1$ and $-2$, for example, under the $U(1)'$ gauge symmetry. After $\phi$ gains its VEV the $U(1)'$ symmetry is broken down to a remnant $\mathbb{Z}_2$ symmetry that stabilises the dark matter\footnote{This model was previously considered in \cite{Ko:2014eqa} with the $U(1)'$ breaking occurring close to the TeV scale.}. At the same time, we generate the $\la_5$ term required with $\la_5\sim v'/\La$.

An alternative is not to charge the inert doublet $H_2$ under the new $U(1)'$ symmetry and we can write down the $\la_5$ term straight away. In this case the $\mathbb{Z}_2$ symmetry required to stabilise our \dm candidate does not have an origin in the broken gauge symmetry, but this is no different from many other \dm models. 
We remain agnostic about the origin of the $\la_5$ term, as we will see both choices lead to the same effective potential. 

After $U(1)'$ symmetry breaking we have two additional massive particles: a $Z'$ gauge boson and a scalar $X$, whose masses are of the order of the symmetry breaking scale $v'$. Since this is far larger than the electroweak scale, we can integrate out these heavy states, yielding
\beq
\begin{split}
V (H_1,H_2) = & -\left(\mu_1^2 - \frac{\tilde{\lambda}_1 v'^2}{2}\right) |H_1|^2 + \left(\mu_2^2 + \frac{\tilde{\lambda}_2 v'^2}{2}\right) |H_2|^2 + \left(\lambda_1 -\frac{\tilde{\lambda}_1^2}{8\lambda_{\phi}}\right) |H_1|^4 \\& + \left(\lambda_2 -\frac{\tilde{\lambda}_2^2}{8\lambda_{\phi}}\right) |H_2|^4 + \left(\lambda_3 -\frac{\tilde{\lambda}_1\tilde{\lambda}_2}{8\lambda_{\phi}}\right) |H_1|^2|H_2|^2 + \lambda_4 |H_1^{\dagger}H_2|^2 \\& + \left\lbrace \frac{\lambda_5}{2}(H_1^{\dagger} H_2)^2 + h.c.\right\rbrace \,.
\end{split}
\eeq
This takes the form (\ref{Potential}) upon the redefinitions
\beq
\label{redef}
\begin{split}
\mu_1^2 - \frac{\tilde{\lambda}_1 v'^2}{2} \to \mu_1^2, \quad \quad &\mu_2^2 + \frac{\tilde{\lambda}_2 v'^2 }{2} \to \mu_2^2, \quad \quad \lambda_1 -\frac{\tilde{\lambda}_1^2}{8\lambda_{\phi}} \to \lambda_1, \\
\lambda_2 -\frac{\tilde{\lambda}_2^2}{8\lambda_{\phi}} \to \lambda_2, \quad \quad &\lambda_3 -\frac{\tilde{\lambda}_1\tilde{\lambda}_2}{8\lambda_{\phi}} \to \lambda_3.
\end{split}
\eeq
We can then consider our effective theory at the electroweak scale to be the standard IDM, with the additional presence of cosmic strings, which couple to dark matter (and the Higgs). This is the case regardless of whether $H_2$ is charged or not under the $U(1)'$ symmetry. 

We note that the first two redefinitions in (\ref{redef}) introduce a large fine-tuning in order for $\mu_{1,2}^2$ to be close to the EW scale. This is of course one of a number of large fine tunings in models of this type. One can think of this as a sensitivity to high scale physics, which we can recognise as the gauge hierarchy problem of the SM. We do not address this issue in this work but a possible solution is to supersymmetrise the model. This is well beyond the scope of this work and we pragmatically regarded this work as a case study of the possible impact of \dm production via TDs in a simple \dm model.

The IDM potential (\ref{Potential}) can be fully described after electroweak symmetry breaking by the parameters $v$, $m_h$, $\lambda_2$, $m_{H^0}$, $\Delta m_0^2$, $\Delta m_+^2$ and $\lambda_L$. The parameters $v$ and $m_h$ are fixed, while $\lambda_2$ does not enter (at tree-level) any of the processes we will consider. Furthermore, to make results more comprehensible (and to be compatible with EWPO), 
we will consider the case where $\Delta m_0^2 = \Delta m_+^2 \equiv \Delta m^2$, which leaves us with three relevant parameters for our analysis: $m_{H^0}$, $\Delta m^2$ and $\lambda_L$. In the next section, we parameterise the TD sector of the model.

\section{Spontaneous $U(1)'$ breaking and cosmic strings}\label{CSSection}
%%%%%%%%%%%%%%%%%%%%%%%%%%%%%%%%%%%%%%%%%%%%%%%%%%%%%%%%

Symmetry-breaking phase transitions in the early universe can give rise to TDs \cite{Kibble:1976sj}, and in the case of a broken $U(1)'$, cosmic strings \cite{Hindmarsh:1994re,VilShe94,Hindmarsh:2011qj} are formed. The scale of the symmetry-breaking, $v'$ dictates the string tension ($\mu \sim {v'}^2$) and the width ($\sim 1/v'$). In the core of the string there is a local false vacuum, where $v'$ vanishes, and if another field couples appropriately to the field breaking the symmetry, it forms a condensate on the string \cite{Witten:1984eb}. In our IDM with an additional $U(1)'$ gauge symmetry, the string is made from the $\phi$ field and the $U(1)'$ gauge field, and can develop condensates of both Higgs doublets, whose amplitude can be as large as $v'$ \cite{Vachaspati:2009kq,Hyde:2013fia,Mota:2014uka}.

As the universe evolves, the cosmic strings decay and produce particles, with an energy density injection rate $Q(t)$. There is also a decay channel into gravitational radiation, which may dominate at late times (for a discussion see \cite{Hindmarsh:2011qj}). We parameterise the time-dependence by \cite{Bhattacharjee:1998qc}
\beq
\label{e:DefInjRat}
Q(t) = Q_\chi\left(\frac{t}{t_\chi}\right)^{p-4},
\eeq
where $t_\chi$ is a reference time in the radiation-dominated universe, set to be when the temperature is equal to the mass of the DM particle. The value of $Q_\chi$ depends on the model parameters $\la_\phi$, $\tilde{\la}_1$, $\tilde{\la}_2$, and most importantly $\mu_\phi$: for details see Appendix  \ref{GmuExpressions} and Ref.\ (\cite{Hindmarsh:2013jha}).
As was shown in the preceding section, these parameters do not appear in the effective EW scale theory, and are therefore very weakly constrained. We will treat $Q_\chi$ as a free parameter to be bounded by cosmological data.

We consider two decay scenarios for dark matter production  \cite{Hindmarsh:2013jha}. 
The first is driven by numerical simulations of the Abelian Higgs model \cite{Vincent:1997cx,Bevis:2006mj,Hindmarsh:2008dw} (the field theory or FT scenario), where particle radiation is dominant, 
and $p=1$ in this case. In the second scenario gravitational radiation from oscillating loops dominates the energy loss, but cusps (a section of the string which has doubled back on itself) allow subdominant string decays into particle radiation \cite{Vilenkin:1981bx} (the cusp emission or CE scenario). This gives $p=7/6$ in the radiation era. In both cases we suppose that a fraction $f_i$ of the energy loss is into particles $i=H_0, A_0,...$, which have an average energy of $\bar{E}_i$.

In the IDM model, where there are condensates of both Higgs doublets, we expect the production of dark matter states to be accompanied by the production of ordinary Higgs states.
Higgs production by strings can have observable consequences \cite{Vachaspati:2009kq,Hyde:2013fia,Mota:2014uka,Long:2014lxa}, which constrains the dark matter injection rate. For example, decays of string-produced Higgs particles can affect the light element abundances during Big-Bang-Nucleosynthesis (BBN), and can also produce photons that will contribute to the Diffuse Gamma-Ray Background (DGRB). Constraints on the energy injection rate into ``visible'' (i.e. electromagnetically interacting) particles from strings were derived in \cite{Mota:2014uka} (see also \cite{Long:2014lxa}).
The effects of these constraints on the available parameter space of the IDM model are discussed in more detail in Section \ref{BBNcons}.

\section{Increasing the relic density in the Inert Doublet Model}\label{sec:boltzsolve}
%%%%%%%%%%%%%%%%%%%%%%%%%%%%%%%%%%%%%%%%%%%%%%%%%%%%%%%

In this section we construct the Boltzmann equations for the production of \dm in our scenario.  The effect of the TD decays is to introduce a source of \dm states with an injection rate parametrised by Equation~\ref{e:DefInjRat}. This type of term was incorporated into Boltzmann equations for a single species of \dm in \cite{Hindmarsh:2013jha}.
In addition to the source term, we will investigate a particular scenario of the IDM where co-annihilations may be important in determining the final relic abundance of \dm states. By generalising the treatments outlined in \cite{Griest:1990kh,Edsjo:1997bg} to include source terms the Boltzmann transport equations, detailing the evolution of the number density $n_i$ for any odd sector particle, can be written as
\beq
\begin{split}
\frac{dn_i}{dt} = & - 3 H n_i -\sum_{j} \langle\si_{ij} v_{ij}\rangle(n_i n_j - n_i^{\text{eq}}n_j^{\text{eq}}) \\& -\sum_{j\neq i}\left[\langle\si'_{ij} v_{ij}\rangle (n_i n_X - n_i^{\text{eq}} n_X^{\text{eq}}) - \langle\si'_{ji} v_{ij}\rangle (n_j n_{X^{\prime}} - n_j^{\text{eq}} n_{X^{\prime}}^{\text{eq}})\right] \\& - \sum_{j \neq i}\left[\Ga_{ij}(n_i - n_i^{\text{eq}}) - \Ga_{ji}(n_j - n_j^{\text{eq}})\right] + \frac{f_i Q(t)}{\Ebar_i},
\end{split}
\label{eq:BE}
\eeq
where $n^{\text{eq}}_i$ is equilibrium number density for particle species $i$. Here, both $X$ and $X^{\prime}$ represent states that are even under the stabilising symmetry $\mathbb{Z}_2$, namely the SM states. The first term on the right hand side of Equation~\ref{eq:BE} accounts for the dilution of the particle number density due to cosmological expansion. The second and third terms take into account annihilation/creation ($ij\leftrightarrow X$) and scattering ($iX \rightarrow jX^{\prime}$) processes respectively, while the fourth takes account of particle decays. The final term represents the contribution from TD decays. The cross sections and decay rates are schematically
\beq
\si_{ij} = \sum_X \si(ij \to X), \quad\quad\quad \si'_{ij} = \sum_{X, X^{\prime}} \si(iX \to jX^{\prime}), \quad\quad\quad
\Ga_{ij} = \sum_X \Ga(i \to jX).
\eeq
To deal with the co-annihilations we can sum the Boltzmann equations for all the odd sector particles, the resulting expression describes the number density evolution of the total odd sector particles ($n = \sum_i n_i$). This procedure removes the scattering and decay terms, as they do not change the overall number of odd particles and as a result cancel amongst each other in the sum. Given that only the lightest of these odd particles is stable, we expect to be left with just the Lightest Odd sector Particle (LOP) at late times (that is, $n(t_0) = n_{\text{LOP}}(t_0)$). The evolution of $n(t)$ can then be written as \cite{Edsjo:1997bg}
\beq\label{TotalBE}
\frac{dn}{dt} = -3Hn - \langle\si_{\text{eff}}v\rangle(n^2 - (n^{\text{eq}})^2) + \sum_i \frac{f_i Q(t)}{\Ebar_i},
\eeq
where
\beq
\langle\si_{\text{eff}} v\rangle \equiv \sum_{ij}\langle\si_{ij}v_{ij}\rangle\frac{n_i^{\text{eq}} n_j^{\text{eq}}}{n^{\text{eq}} n^{\text{eq}}}.
\eeq
Performing the standard manipulations; converting from time variable $t$ to the variable $x\equiv m_{H^0}/T$ and replacing the number density with $n=Ys$, where $s$ is the entropy density and $Y$ is the yield, the resulting form of the Boltzmann equation can then be written in the simplified form
\beq
\frac{dY}{dx} = -\frac{A(x)}{x^2} \left[Y^2 - (Y^{\text{eq}})^2\right] + \frac{B}{x^{4-2p}},
\eeq
where
\beq
A(x) = \sqrt{\frac{\pi g_*}{45}}m_{H^0}M_{\text{Pl}} \langle\si_{\text{eff}} v\rangle(x), \qquad B = \frac{3}{4} r_0 q_0,
\eeq
with
\beq\label{q0Def}
q_0 = \sum_i \frac{\Ebar_{H^0}}{\Ebar_i}\frac{f_i Q_\chi}{\rho_\chi H_\chi},\;\;\mbox{and}\;\;r_0 = \mH/\Ebar_{H^0}.
\eeq
The equilibrium yield takes the form
\beq
\label{FinalBE}
Y^{\text{eq}} = \frac{45 x^2}{4\pi^4 h_{\text{eff}}}\sum_i g_i\left(\frac{ m_i}{m_{H^0}}\right)^2 K_2\left(x\frac{m_i}{m_{H^0}}\right).
\eeq
In the above expressions $\Mpl = 1.22 \times 10^{19}$ GeV, $g_i$ is the degrees of freedom of particle $i$, and $\rho_\chi$ and $H_\chi$ are the energy density and Hubble parameter values at $t_\chi$, respectively. 
We use
\bea
\sqrt{g_*}= \frac{h_{\text{eff}}}{g_*^{\rho}}\left(1+\frac{T}{3h_{\text{eff}}}\frac{dh_{\text{eff}}}{dT}\right),
\eea
where $g_*^{\rho}$ and $h_{\text{eff}}$ are the effective numbers of degrees of freedom in the bath for the energy density and entropy respectively.
We set the parameter $r_0=0.5$ in the following analysis, which is typical for strings with condensates of light fields \cite{Vachaspati:2009kq}. The $q_0$ parameter encodes the total energy injection rate in the form of odd sector particles from the TD decays, its value is almost entirely set by the symmetry breaking scale, with some weak dependence on other variables, see Appendix \ref{GmuExpressions}. For sufficient dark matter production $q_0$ is typically required to be larger than $\sim 10^{-12}$, corresponding to a minimum $v'$ value of $\sim 10^{11}$ GeV.

\subsection{Numerical evaluation of the relic abundance}

Throughout our study we focus on four benchmark points for the two TD decay scenarios of $p=1, 7/6$. These four benchmarks are determined by two values of $\De m^2$ and two values of the coupling $\la_L $, namely, $\De m^2 = 1000, 10000$\;GeV$^2$ and $\la_L = 0, 0.1$.

The two choices of mass squared splittings correspond to cases where co-annihilations are and are not important, respectively. The value of $\la_L = 0$ is a special limiting case for which direct detection signals are absent at leading order, as the $H^0 H^0 h$ coupling is proportional to $\la_L$. In addition, $\la_L = 0$ also suppresses the annihilation $H^0 H^0 \to \bar{f}f$ which is usually the main annihilation channel in the low mass ($m_{H^0} < m_W$) region.
\begin{figure}[t]
\begin{center}
\hspace{-0mm}\includegraphics[width=0.48\textwidth]{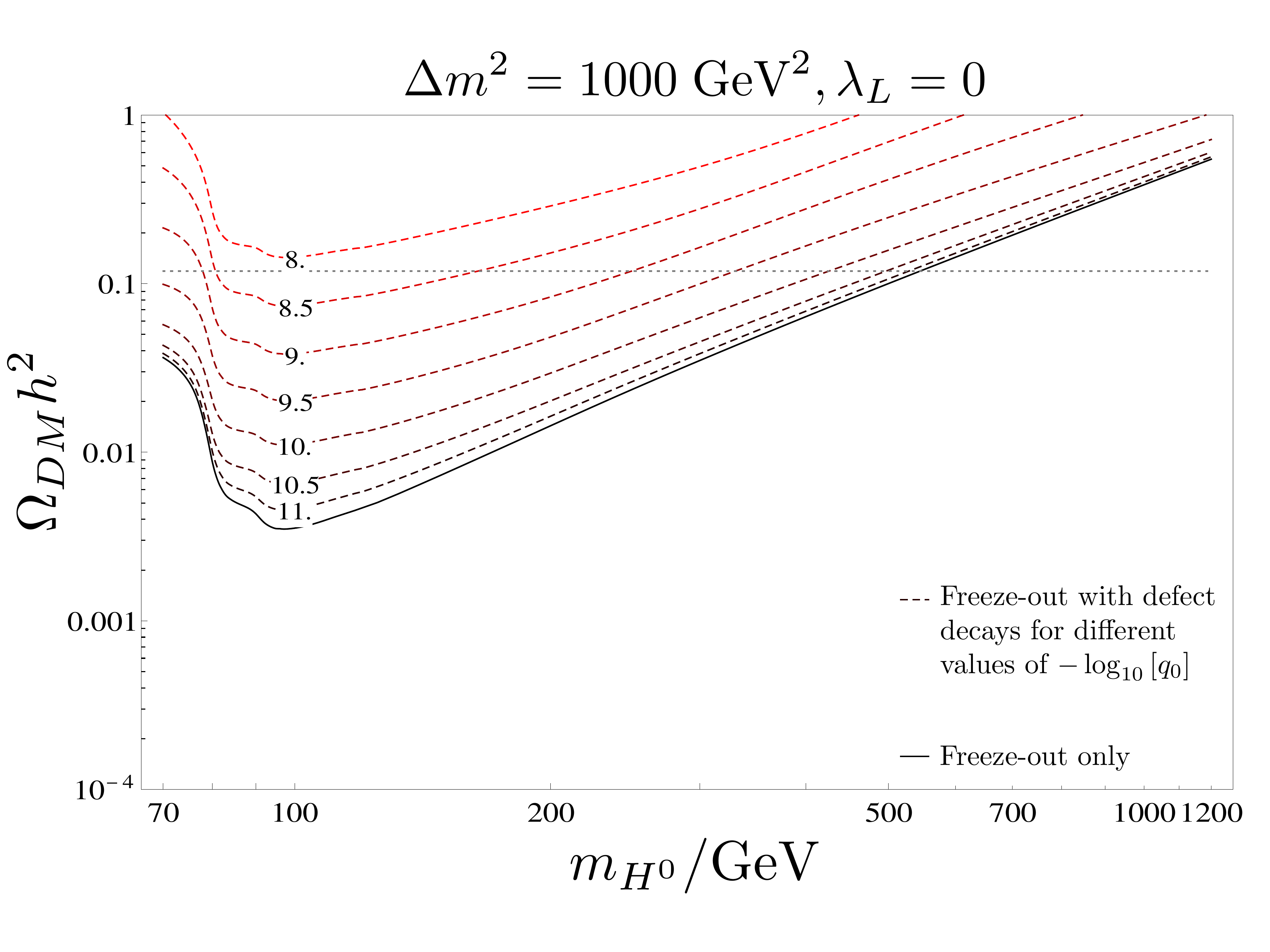} \hspace{3mm}
\includegraphics[width=0.48\textwidth]{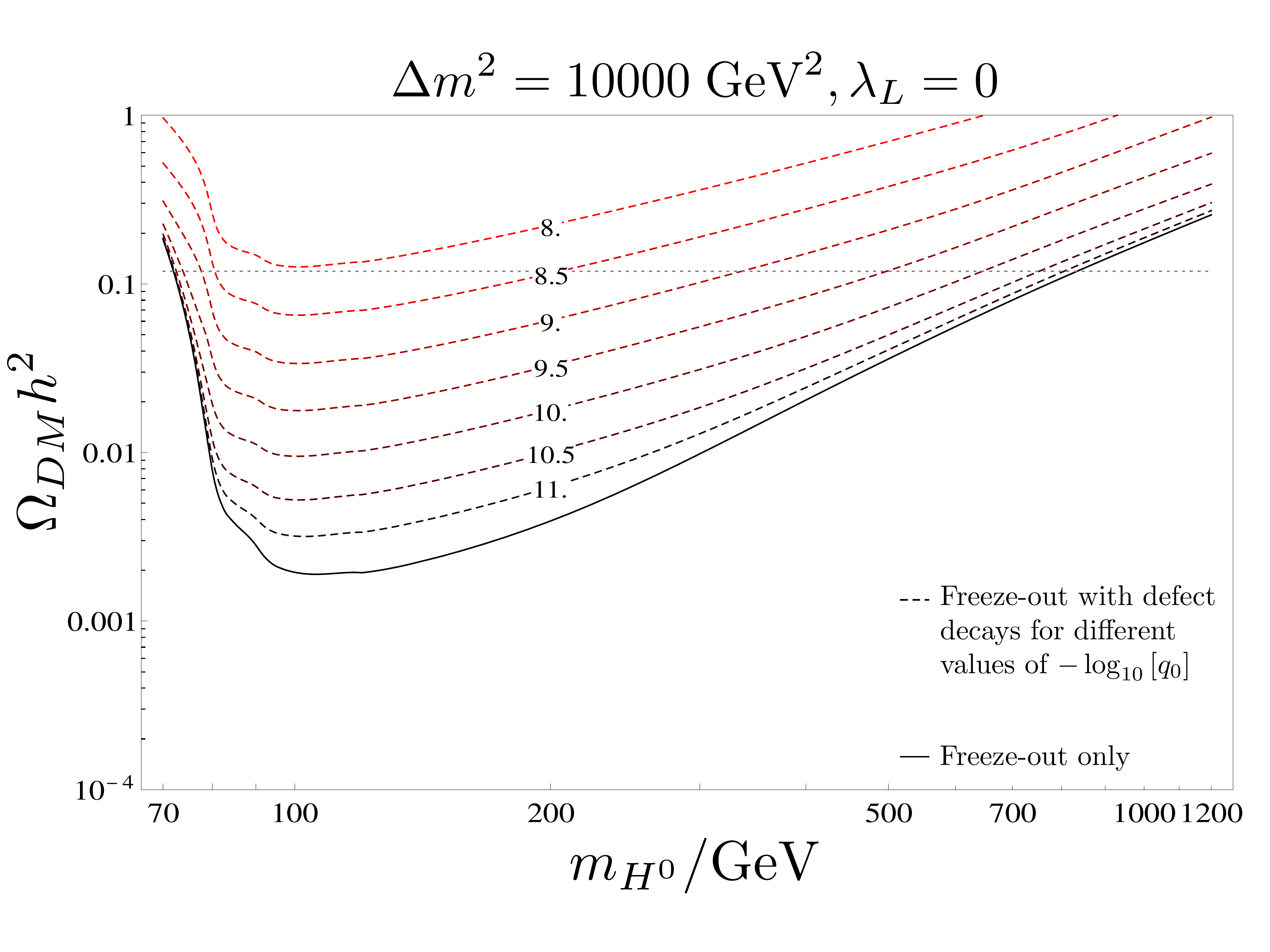}
\includegraphics[width=0.48\textwidth]{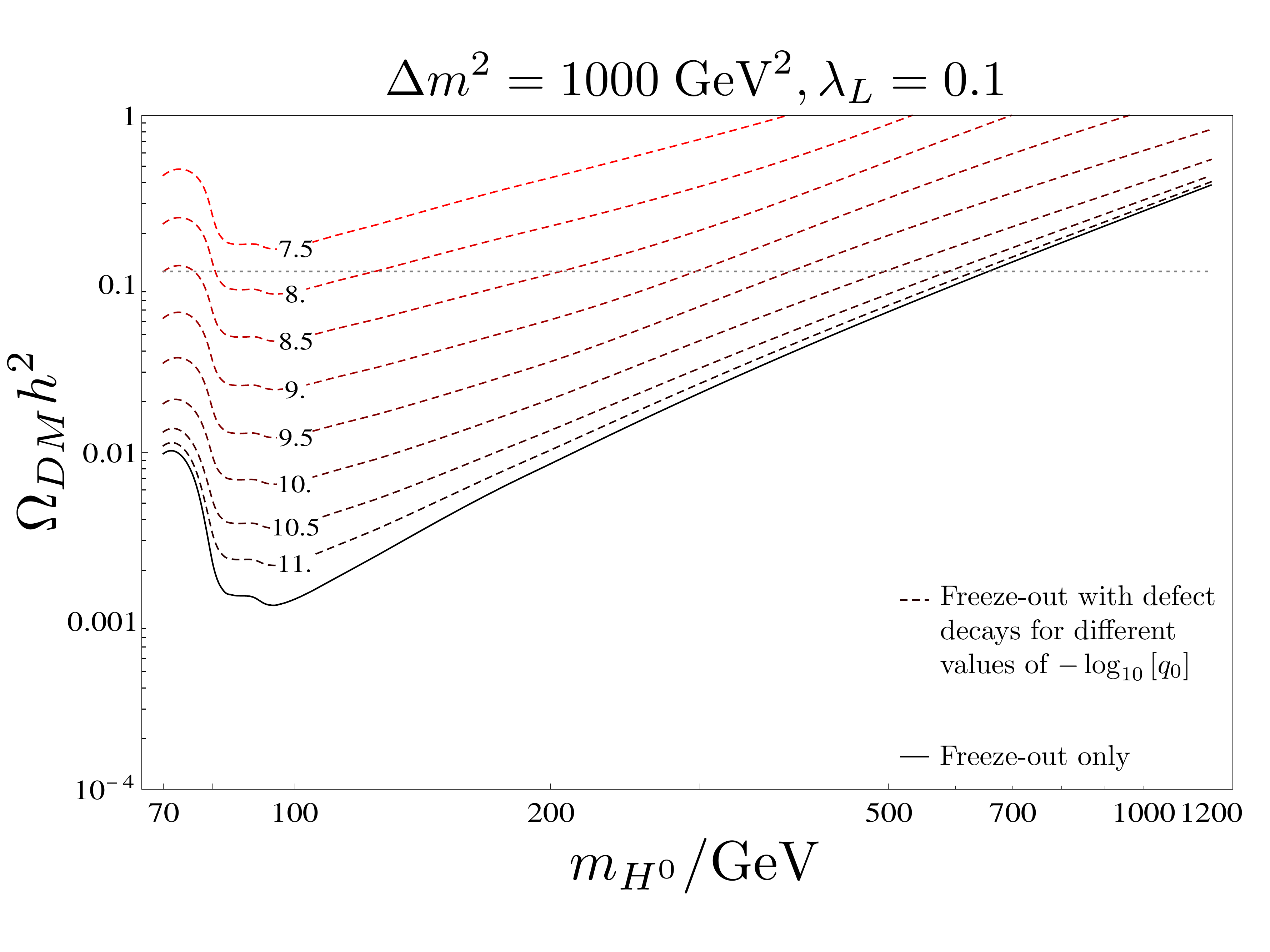}\hspace{4.4mm}
\includegraphics[width=0.48\textwidth]{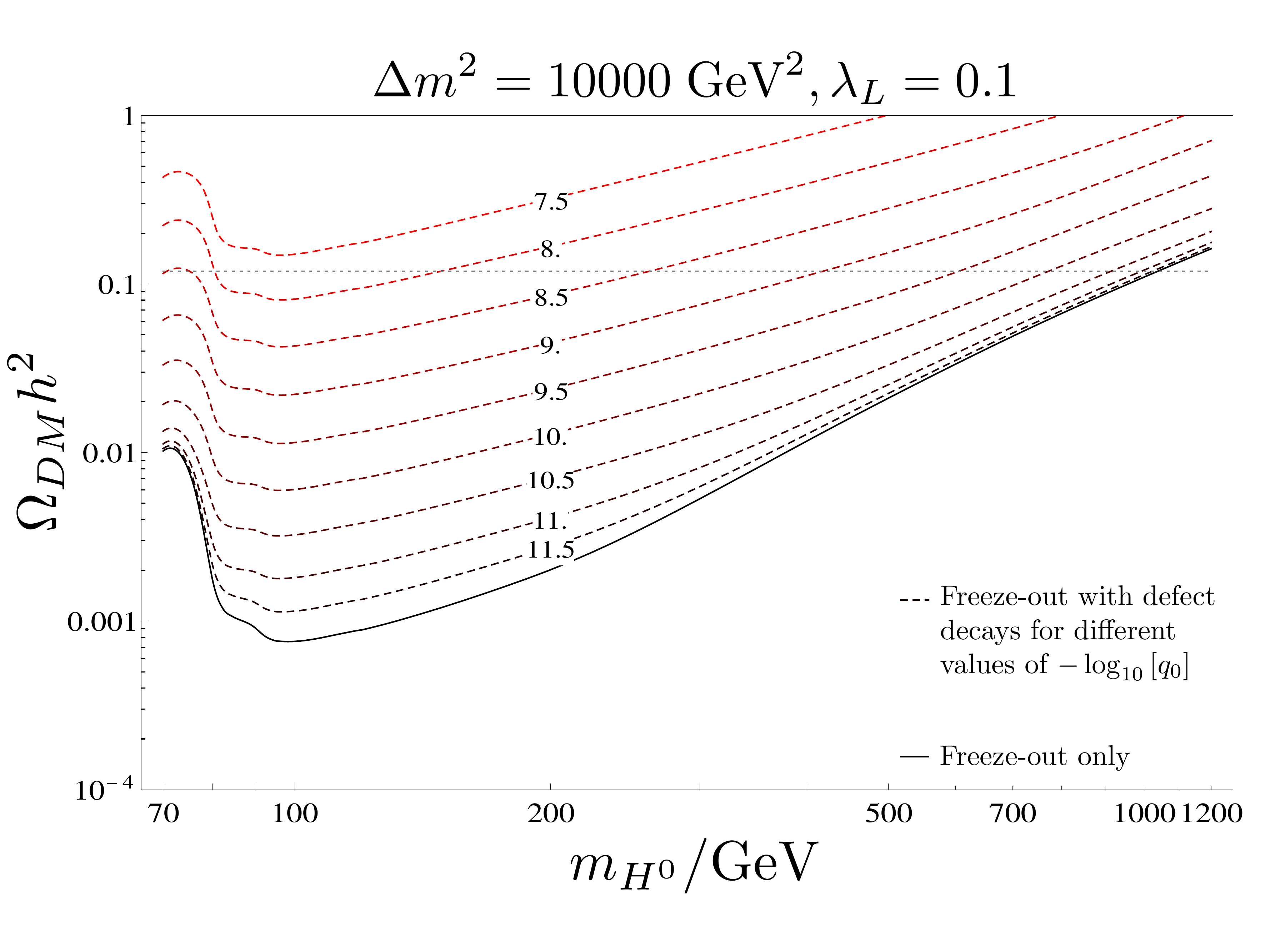}
\caption{Relic abundance vs $m_{H^0}$ for $(\De m^2, \la_L)$ $ = (1000\, \text{GeV}^2, 0)$, $(10000\, \text{GeV}^2, 0)$, 
$(1000\, \text{GeV}^2, 0.1)$, $(10000\, \text{GeV}^2, 0.1)$. The solid-black line corresponds to standard thermal freeze-out, while the dotted lines correspond to scenarios with contributions from defect decays with varying $-\log_{10}(q_0)$ values (FT scenario, $p=1$).}
\label{OmegaPlots}
\end{center}
\end{figure}

In order to solve the Boltzmann equation with these parameters we have used micrOMEGAS \cite{Belanger:2013oya} to evaluate $\langle\si_{\text{eff}} v \rangle$.
The results are plotted in Figure~\ref{OmegaPlots}, which shows the variation of the predicted relic abundance as a function of \dm mass for a range of $q_0$ values for the FT ($p=1$) case. We have chosen a minimum mass of 70\;GeV, as below this mass the collider constraints on the model become quite restrictive. 

The corresponding plots for the CE ($p=7/6$) case are very similar with a slight shift down in the value of $q_0$ needed to attain the correct relic abundance. With a larger power $p$ the TD will have a larger injection rate and will therefore require a smaller $q_0$ to generate the same relic abundance. We do not show these plots as apart from this slight shift they are very similar to the FT plots. 

Shown in each plot as solid black contours is the freeze-out only case with $q_0$ set to zero. This, as we would expect, generates the lowest value of the relic abundance for a given mass within each benchmark point. As the value of $q_0$ is increased, the predicted value of the relic abundance also increases. We note that in each case, the freeze-out only scenario can generate the correct relic abundance for just one mass value. Below this mass, the freeze-out only case predicts an abundance that is too small to account for all of the dark matter. It is apparent that for each mass considered there is a value of $q_0$ for which the observed relic abundance is obtained up to the mass where standard freeze-out is itself over-producing dark matter. All points survive the constraints imposed by perturbativity, vacuum stability and unitarity constraints as discussed in Section \ref{IDMReview}. 

As mentioned above, a consequence of small $\De m^2$ values is the increase of co-annihilations in the freeze-out process. Generally, the greater the importance of co-annihilations, the smaller the relic abundance will be. However, as we can see from Figure~\ref{OmegaPlots}, the opposite effect is observed in the IDM: comparing the $\De m^2=1000\;$GeV$^2$ plots with the ones for $\De m^2=10000\;$GeV$^2$ we see that in fact the smaller mass splitting generates the larger abundance, despite the greater contribution of co-annihilations. 

To understand the physical reason behind this we first note that for 
$m_{H^0} > m_V$ ($V = W^{\pm},\, Z$) dark matter annihilation into gauge bosons $H^0\,H^0 \to V\, V$ is the dominant annihilation process, and generally yields a relic density significantly below the observed one\footnote{This is a generic feature of models in which dark matter annihilates via gauge interactions, {\it \'a la}  Minimal Dark Matter \cite{Cirelli:2005uq}.} for 
$m_W \ll m_{H^0} \lesssim 1\,\,\mathrm{TeV}$ \cite{Barbieri:2006dq,LopezHonorez:2006gr}. 
However, in the limit $\De m^2 = 0$ (which, together with $\lambda_L = 0$, is the ``pure gauge" limit discussed in \cite{Hambye:2009pw}) there is a cancellation 
among the various terms in the expansion of the annihilation amplitude in powers of $m^2_{H^0}/m^2_{W}$.  

The underlying reason is gauge invariance, which ensures that the amplitude squared remains unitary for high \dm masses. As $\De m^2$ grows, this unitarity cancellation ceases to be exact, and the net result is an increase the effective annihilation cross section $\langle\si_{\text{eff}} v \rangle$. A more detailed discussion may be found in Appendix \ref{AppA2}.

Note that the plots in Figure \ref{OmegaPlots} do not include the experimental constraints on the various IDM parameters coming from \dm phenomenology. Neither do they include constraints from cosmological bounds on $q_0$ from BBN and DGRB. These constraints, and their impact on the allowed parameter space of the model, are analysed in the next section.

\section{Constraints on the Inert Doublet Model with topological defects}\label{sec:DefectConstraints}
%%%%%%%%%%%%%%%%%%%%%%%%%%%%%%%%%%%%%%%%%%%%%%%%%%%%%%%%

In this section we go through the principal observational constraints on the model, in particular those coming from direct and indirect \dm detection experiments, and those coming from BBN and the DGRB.

\subsection{Direct detection}

The most stringent experimental bounds on dark matter direct detection for the $m_{H^0}$ range we consider currently come from the Large Underground Xenon (LUX) experiment at the Sanford Underground Research Laboratory, which looks for \dm scattering off $118$ kgs (the fiducial target mass) of Xenon. In the IDM, dark matter direct detection can occur via a $t$-channel exchange of a SM Higgs between $H^0$ and the nucleon, with amplitude proportional to $\la_L$. Limits on the WIMP-nucleon cross section from $85.3$ live-days of LUX data are presented in \cite{Akerib:2013tjd}, which we implement in our analysis using cross sections calculated in micrOMEGAS \cite{Belanger:2013oya}.

\subsection{Indirect detection}

The Large Area Telescope on the Fermi Gamma-Ray Space Telescope (Fermi-LAT) has mapped the sky in $\ga$-rays in the $20$ MeV to $300$ GeV energy range. Dark matter annihilations can produce such $\ga$-rays and thus Fermi-LAT data may be used to limit the annihilation cross section. The gamma-ray flux is calculated using
\beq
\Phi = \Phi_{\text{PP}} \times J,
\eeq
where
\beq
\Phi_{\text{PP}} = \frac{\langle\si v\rangle}{8\pi m_{H_0}^2}\int_{E_0}^{E_{\text{max}}} \frac{dN}{dE}dE, \quad\quad J = \underset{\De \Om(\psi)}{\int}\underset{l}{\int} [\rho(l,\psi)]^2dl d\Om(\psi).
\eeq
Here $\Phi_{\text{PP}}$ is referred to as the particle physics input, depending on the dark matter annihilation cross section and the photon spectrum produced by the model, $J$ is the astrophysical input, which depends on the local \dm density profile of the chosen celestial body. There is significant uncertainty in this quantity, which produces an uncertainty in the photon flux. We use a combined analysis of the continuum in the range $1 - 100$ GeV from several dwarf spheroidal satellite galaxies (dSphs) \cite{GeringerSameth:2011iw} to constrain the model. This analysis weights the dSphs by their $J$ values and produces a 95\% CL limit of $\Phi_{\text{PP}} < 5.0^{+4.3}_{-4.5}\times 10^{-30} \text{cm}^3 \text{s}^{-1} \text{GeV}^{-2}$. We use micrOMEGAS to calculate the predicted $\Phi_{\text{PP}}$ values for points in our parameter space, then use the central value from the combined dSphs analysis to constrain the space.

Dark matter annihilations can also affect the process of recombination via ionisation and reheating, thus from CMB data one can extract limits on $\langle \si v \rangle$ (see \cite{Galli:2009zc} and references therein). We checked the constraints from WMAP 5-year data, and found that they were weaker than the LUX limits. The projected constraints from Planck data are expected to be competitive with LUX, and possibly superior at low mass.

\subsection{Collider bounds}

Collider experiments can also impose limits on the IDM in a variety of ways. We have already considered constraints from EWPO in Section \ref{EWPOsec} and have shown that for our choice of benchmarks they make no restrictions. For the range of \dm masses we are considering, 
bounds on the invisible decay width of the Higgs do not apply, and LEP bounds on the masses $m_{A^0}$ and $m_{H^\pm}$ as a function of $m_{H^0}$ \cite{Lundstrom:2008ai,Pierce:2007ut} are only relevant for $m_{H^0} < m_W$, which is outside of the range we consider here. We therefore focus in the following on limits from LHC mono-jet searches and on constraints from the Higgs signal strengths measured in the di-photon decay channel.

\subsubsection{Mono-jet (and other) searches at LHC}

The production of $H^0 H^0 \,j$ at the LHC is mediated at tree-level by a SM Higgs. Just as for the case of dark matter direct detection, the signal is suppressed for $\la_L \to 0$. For $\la_L \neq 0$, we can 
extract limits on the IDM parameter space from LHC mono-jet searches using the simplified models analysed in \cite{deSimone:2014pda}, which contain a scalar dark matter candidate coupled to the SM Higgs. Using $\sqrt{s} = 8$ TeV LHC data, the analysis from \cite{deSimone:2014pda} finds a limit $m_{H^0} \gtrsim m_h/2$ at 90\% C.L., which is automatically satisfied in our analysis due to our choice of mass range for $H^0$. 

\vspace{1mm}

In the limit $\Delta m^2 \to 0$, $A^0 H^0 \,j$ production at the LHC mediated by a $Z$ boson will also give rise to a mono-jet signal, since the visible decay products of $A_0$ are expected to be very soft, below the ATLAS and CMS trigger thresholds. In this case the limits from the simplified model in \cite{deSimone:2014pda} with dark matter coupled to a $Z$ boson apply, which however only constrain
a coupling $g_{A^0 H^0 Z} \lsim 5$ for $m_{H^0} > m_W$. These limits are then easily avoided in our set-up.

Finally, we also comment on the case of a mass splitting $\Delta m^2$ large enough for the visible decay products of $A_0$ and $H^\pm$ to be detected by ATLAS and CMS. In this case, the relevant LHC search is analogous to that of chargino/neutralino pair production in supersymmetric models, leading to multilepton signatures and missing transverse energy. We note that the latest ATLAS results 
using the full $\sqrt{s} = 8$ TeV dataset \cite{Aad:2014nua} only constrain neutralino masses $\lesssim 100-120$ GeV for the decay pattern relevant to us, and moreover the LHC dark matter production cross sections for the IDM will be smaller due to the scalar nature of $H_2$ as opposed to SUSY charginos/neutralinos. Once more, this does not place any relevant constraint on the IDM parameter space for   
$m_{H^0} > m_W$.

\subsubsection{$h\to \ga\ga$ signal strengths}

Recently both CMS and ATLAS released updated analyses on the Higgs decay to two photons, measuring signal strengths of $R^{\text{CMS}}_{\ga\ga} = 1.14^{+0.26}_{-0.23}$ \cite{Khachatryan:2014ira} and $R^{\text{ATLAS}}_{\ga\ga} = 1.17 \pm 0.27$ \cite{Aad:2014eha}, respectively. In the IDM the charged scalar $H^\pm$ gives additional contributions to $h\to \ga\ga$ decays, via a triangle loop. The di-photon signal strength for a SM Higgs boson in the IDM, $R_{\ga\ga}$, is given by
\beq
R_{\ga\ga} \equiv \frac{\si(pp \to h \to \ga\ga)^{\text{IDM}}}{\si(pp \to h \to \ga\ga)^{\text{SM}}} \approx \frac{\si(gg\to h)^{\text{IDM}} BR(h\to \ga\ga)^{\text{IDM}}}{\si(gg\to h)^{\text{SM}} BR(h\to \ga\ga)^{\text{SM}}} = \frac{BR(h\to \ga\ga)^{\text{IDM}}}{BR(h\to \ga\ga)^{\text{SM}}}.
\eeq
We may approximate $R_{\ga\ga}$ as the ratio of decay rates $\Ga(h \to \ga\ga)$ in the IDM and SM, since the difference in the total Higgs decay width will be negligible for $m_{H^0} > m_h/2$. The analytic expression for $R_{\ga\ga}$ is given in \cite{Swiezewska:2012eh}, which we use to calculate the theoretical signal strength within the parameter space we consider. We find at most a 10\% deviation from the SM result. Thus the predicted contribution from the IDM is consistent with the experimentally measured values quoted above for the parameter space we consider.

\subsection{Big Bang Nucleosynthesis and the diffuse $\ga$-ray background}
\label{BBNcons}

Considering the cosmic string side there are also cosmological bounds on $q_0$ \cite{MacGibbon:1989kk,Sigl:1995kk,Wichoski:1998kh,Mota:2014uka,Long:2014lxa,Berezinsky:2011cp}. Injection of high energy particles after the beginning of nucleosynthesis can alter the abundances of light elements and spoil the agreement of standard BBN with the observed light element abundances, in particular $^4$He and D. Energy injected during recombination produces distortions in the CMB energy spectrum, while energy injected after recombination initiates electromagnetic cascades through collisions with the cosmic medium, and shows up as $\ga$-rays with energies below the threshold for $e^+e^-$ pair production from background electromagnetic radiation. These cascades also produce high energy neutrinos, which can be detected in neutrino telescopes. 

The strongest bounds on the IDM with TDs are from BBN and the DGRB \cite{Mota:2014uka,Long:2014lxa}, which constrain the injection rate into ``visible'' SM particles ($\ga$, $e$, $p$ and $n$). We denote the corresponding energy injection rate as $\QVis$, with a dimensionless version $\qVis$ defined analogously to $q_0$ (Equation \ref{q0Def}). It turns out that BBN constrains the CE scenario ($p=7/6$) most strongly, while the DGRB constraints and BBN constraints from the deuterium abundance are about the same for the FT model ($p=1$).

Let us consider the DGRB constraint on the FT model first. Fermi-LAT data limits the energy injection into visible SM particles to being less than 
$\om_\text{cas}^\text{max} = 5.8 \times 10^{-7} \; \text{eV}\,\text{cm}^{-3}$ \cite{Abdo:2010nz,Berezinsky:2010xa}.
Hence
\ben
\int_{t_c}^{t_0} dt \left( \frac{a(t)}{a(t_0)}\right)^4 \QVis \lesssim \om_\text{cas}^\text{max},
\een
where $t_c \simeq 10^{15}\,\text{s}$ is the time at which the universe became transparent to $\ga$-rays to which Fermi-LAT is sensitive. For $p=1$ it is straightforward to show that
\ben
\qVis \lesssim \left( \frac{t_0}{t_c}\right)^\frac{2}{3} \frac{\om_\text{cas}^\text{max}}{\rho_0},
\een
where $\rho_0\simeq 5.6\times 10^3 \; \text{eV}\,\text{cm}^{-3}$ is the energy density today. Hence
\ben\label{DGRBBound}
\qVis \lesssim 6 \times 10^{-9}  \left( \frac{10^{15}\,\text{s}}{t_c}\right)^\frac{2}{3} .
\een

BBN bounds the energy injected per unit entropy by an $X$ particle with lifetime $\tau_X$, or $E_\text{vis}Y_X$ \cite{Kawasaki:2004qu}. For the $p=7/6$ model the bound with the best combination of strength and robustness comes from the deuterium abundance \cite{Mota:2014uka}, which limits the energy injected per unit entropy to
\ben
(E_\text{vis}Y_X)_\text{max} \simeq 10^{-13}\;\text{GeV},
\een
for $\tau_X \simeq 400\,\text{s}$. We can approximate the energy density injected at time $t$ by $t\QVis$, and hence
\ben
\qVis(\tBBN) \lesssim \frac{8}{3} \frac{(E_\text{vis}Y_X)_\text{max}}{\TBBN},
\een
where $\tBBN \simeq 400\,\text{s}$, and $\TBBN$ is the corresponding temperature.

We must take into account the time-dependence of $\qVis$ in the $p=7/6$ model in order to relate the BBN bound to the energy injection rate at the reference time for dark matter freeze out, or $\tref$. Hence
\ben
\qVis(t_0) \lesssim \frac{8}{3} \frac{(E_\text{vis}Y_X)_\text{max}}{\TBBN} \left( \frac{\tref}{\tBBN} \right)^\frac{1}{6}.
\een
Substituting in the numerical values, and recalling that, at the reference time $\tref$, the temperature is equal to the mass of the dark matter particle $H^0$, we find
\ben\label{BBNBound}
\qVis(t_0) \lesssim 4 \times 10^{-11}  \left( \frac{100\,\text{GeV}}{m_{H^0}} \right)^\frac{1}{3}.
\een

\subsection{Results}

\begin{figure}[t]
\begin{center}
\includegraphics[width=0.5\textwidth]{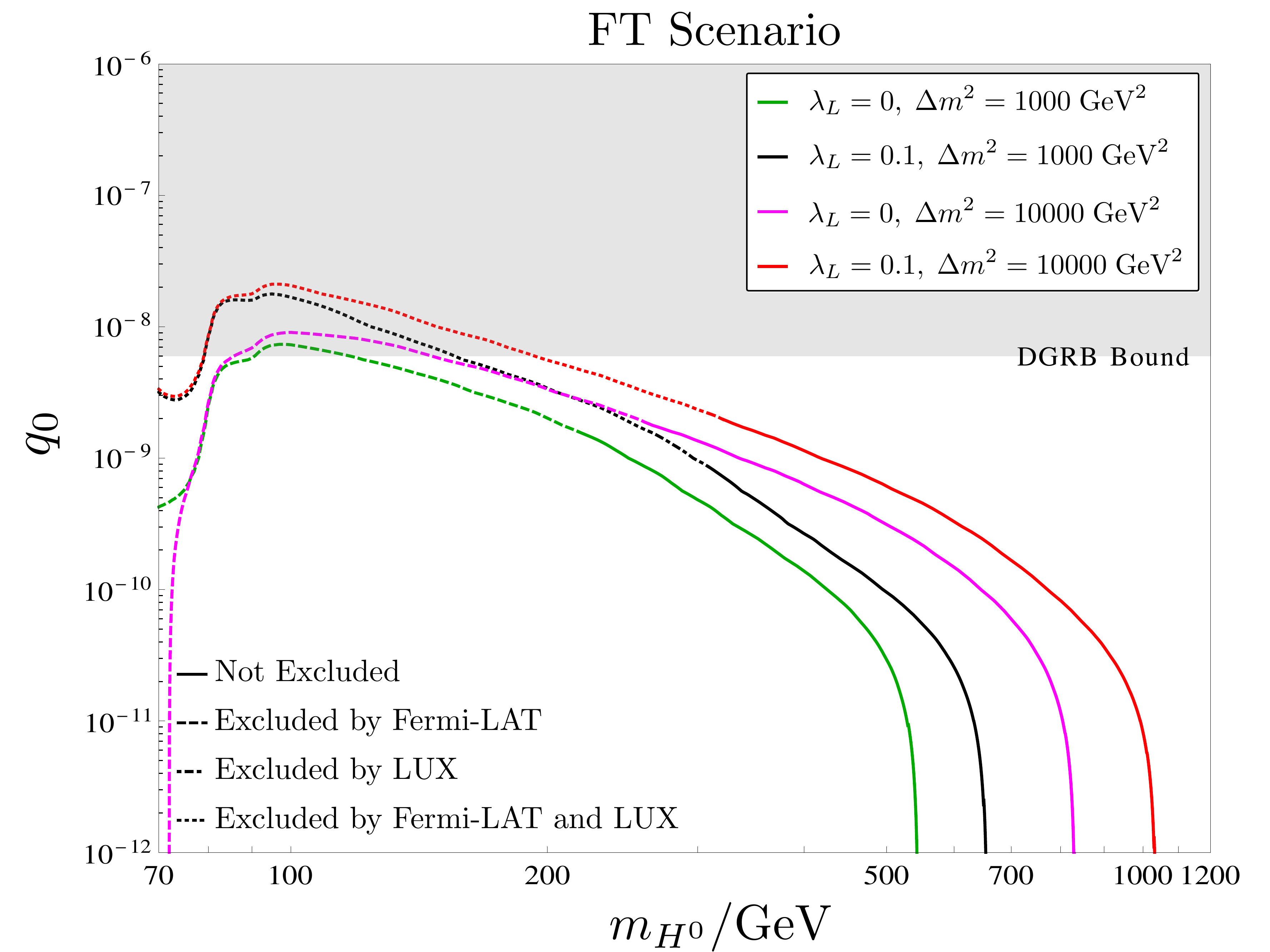}\hspace{-2mm} \includegraphics[width=0.5\textwidth]{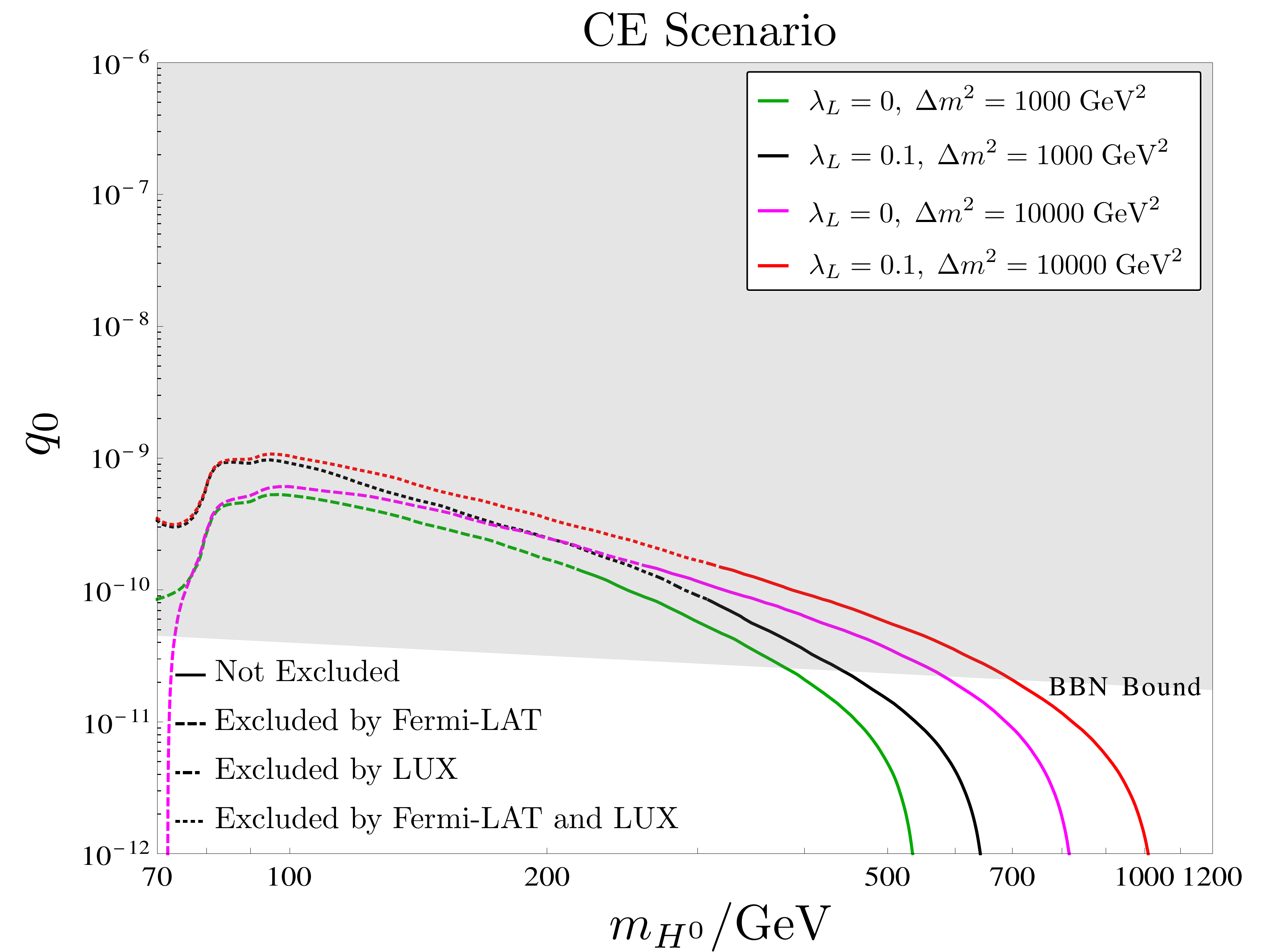}
\caption{For the four benchmark ($\la_L$, $\De m^2$) values in the FT (left) and CE (right) scenarios, the values of $q_0$ required to generate the observed \dm abundance are plotted as a function of $m_{H^0}$. Parameter values that are excluded by LUX and Fermi-LAT are represented by the changes in line style from solid (allowed) to broken (excluded). Regions constrained by the DGRB and BBN (see Section \ref{BBNcons}) are shaded grey.}
\label{q0Plots}
\end{center}
\end{figure}

\begin{table}
\begin{center}
\renewcommand*{\arraystretch}{1.2}
\begin{tabular}{|c|c|c|c|c|}
\hline
\multirow{2}{*}{$(\la_L,\De m^2/\text{GeV}^2)$} &  \multicolumn{2}{|c|}{$m_{H^0}/\text{GeV}$} & \multicolumn{2}{|c|}{$q_0$} \\ \cline{2-5}
 &  FT & CE & FT & CE \\ \hline
 $(0,1000)$ & $220-550$ & $380-550$ & $\lesssim 2\times 10^{-9}$ & $\lesssim 3\times 10^{-11}$ \\ \hline
  $(0.1,1000)$ & $310 - 660$ & $440 - 660$ & $\lesssim 9\times 10^{-10}$ & $\lesssim 2\times 10^{-11}$ \\ \hline
   $(0,10000)$ & $260 - 830$ &  $580 - 830$ & $\lesssim 2\times 10^{-9}$ & $\lesssim 2\times 10^{-11}$ \\ \hline
    $(0.1,10000)$ & $320 - 1040$ & $690 - 1040$ & $\lesssim 2\times 10^{-9}$ & $\lesssim 2\times 10^{-11}$ \\ \hline
\end{tabular}
\end{center}
\caption{Listed are the allowed ranges of $m_{H^0}$ in both FT and CE scenarios for the chosen benchmark $\la_L$, $\De m^2$ values. Also displayed are the absolute upper limits on the values of $q_0$ in each case (corresponding to the value of $q_0$ required to generate the correct relic abundance for the lowest allowed \dm mass).}
\label{q0Limits}
\end{table}

In Figure \ref{q0Plots}, we show the value of $q_0$ required to produce the observed dark matter abundance as a function of the dark matter mass $m_{H^0}$, for the selected benchmark values of the mass squared splitting $\Delta m^2$ and coupling parameter $\la_L$. In each case, the region of parameter space ruled out by Fermi-LAT/LUX experimental data is indicated with broken line styles, while the grey shaded region is ruled out by cosmological bounds on $q_0$ (see Section \ref{BBNcons}). The freeze-out only scenario is also represented in both plots by considering that as we decrease the size of $q_0$ the lines become vertical as they approach the $m_{H^0}$ axis. To a good approximation, the point at which they cross the mass axis is the mass value that gives the correct relic abundance for the particular benchmark in the freeze-out only case. Put another way, for the freeze-out only case, there is one mass value for each benchmark point that generates the required relic abundance. Whereas with the 
addition of the TD decays, there is a range of masses that can generate the correct abundance and so the viable parameter space of the IDM can be expanded compared to the freeze-out only case.

Looking more closely at the FT scenario in Figure~\ref{q0Plots}, we can see that the leading constraint comes from the direct and indirect detection limits rather than the DGRB bound, where as for the CE scenario the range of possible masses is determined by the BBN bound. The allowed range of masses in the FT case is as a result larger. 

Table \ref{q0Limits} lists the ranges of allowed \dm mass for both FT and CE scenarios with the upper allowed value of $q_0$ in each of the benchmark $\la_L$ and $\De m^2$ cases. As indicated in Figure~\ref{q0Plots}, the upper limit on $q_0$ in the FT case comes from the constraints on the resulting properties of the \dm state where as in the CE case the upper limit is derived from the properties of the TDs and their potential impact on BBN. What we discover then, is that the \dm phenomenology in the IDM can only limit the size of $q_0$ in the FT case for these values of $\la_L$. The upper limit in the mass range corresponds to the freeze-out only case. We see clearly that the allowed mass range in each case can be significantly increased with the inclusion of TD decays.

\begin{figure}[t]
\begin{center}
\hspace{-1mm}\includegraphics[width=0.48\textwidth]{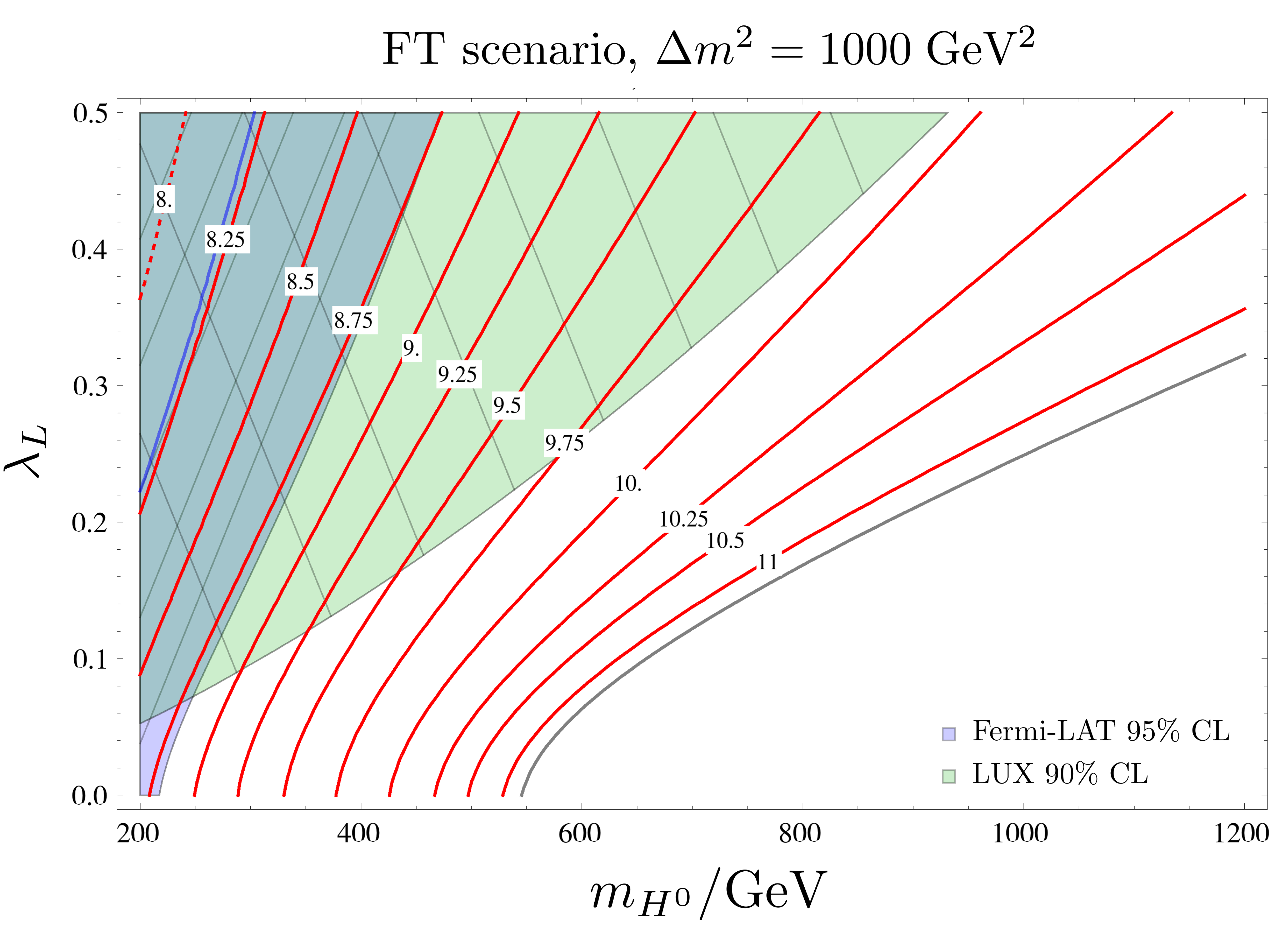}\hspace{3mm} \includegraphics[width=0.48\textwidth]{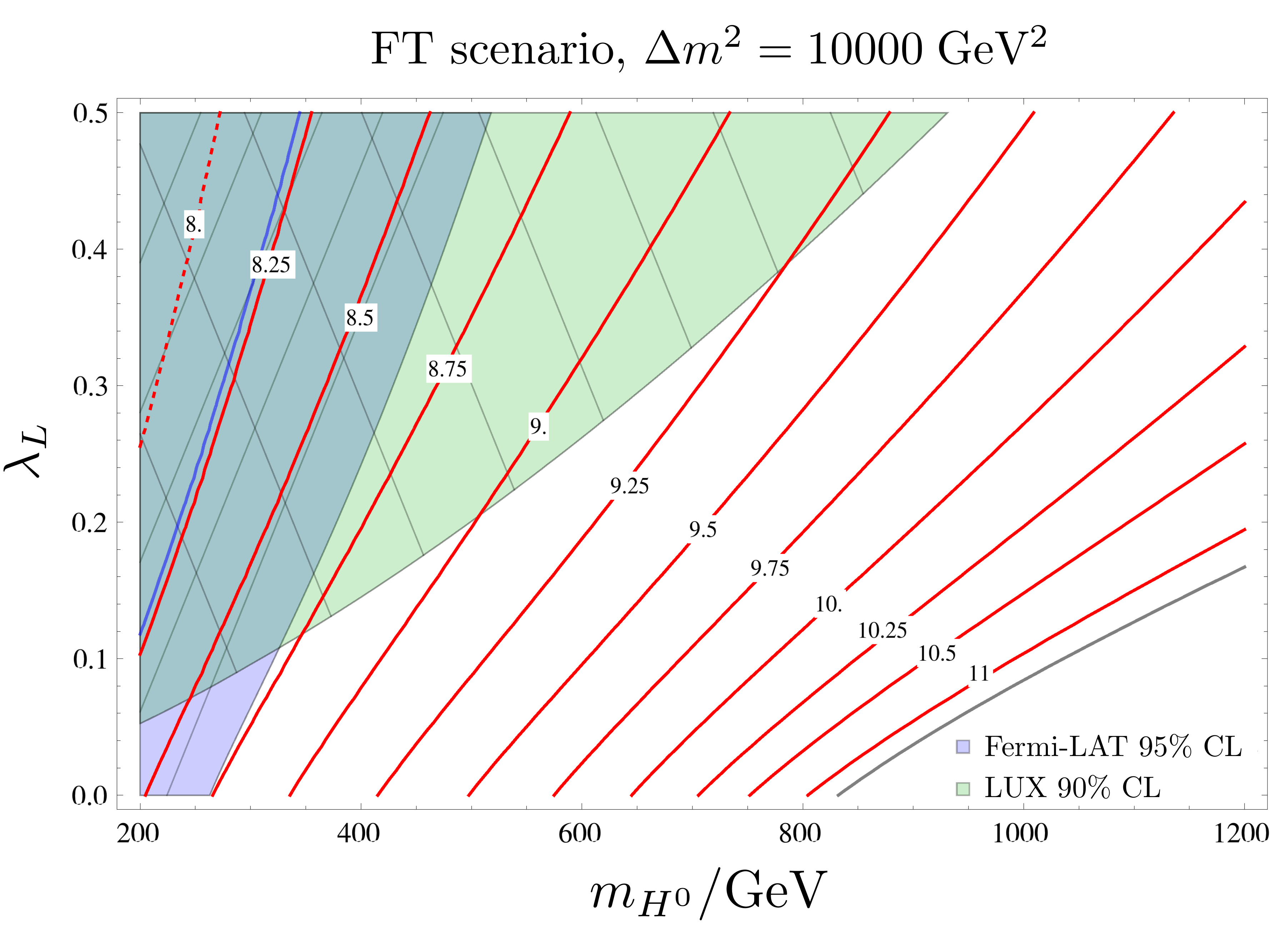}
\includegraphics[width=0.48\textwidth]{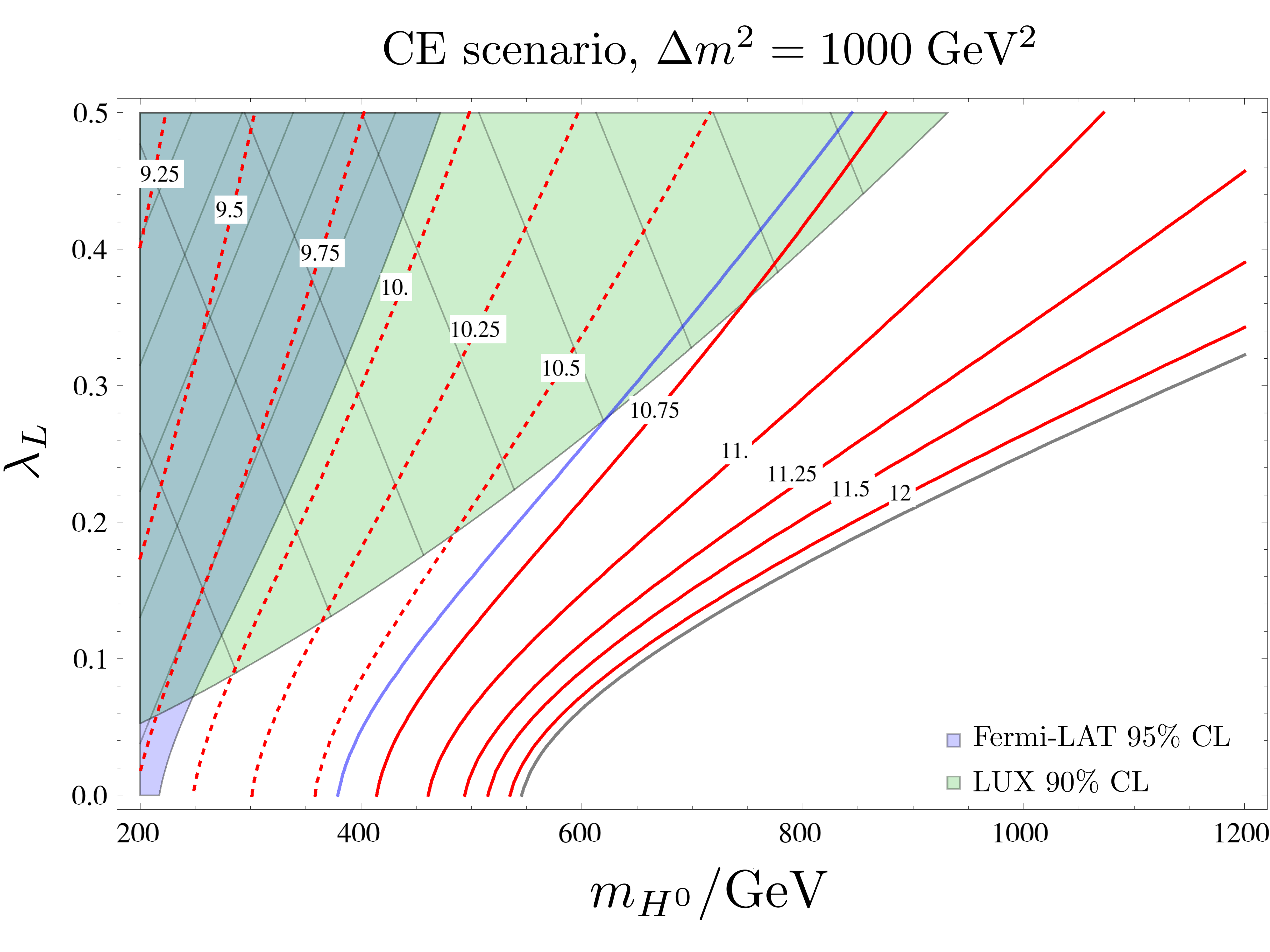}\hspace{3mm} \includegraphics[width=0.48\textwidth]{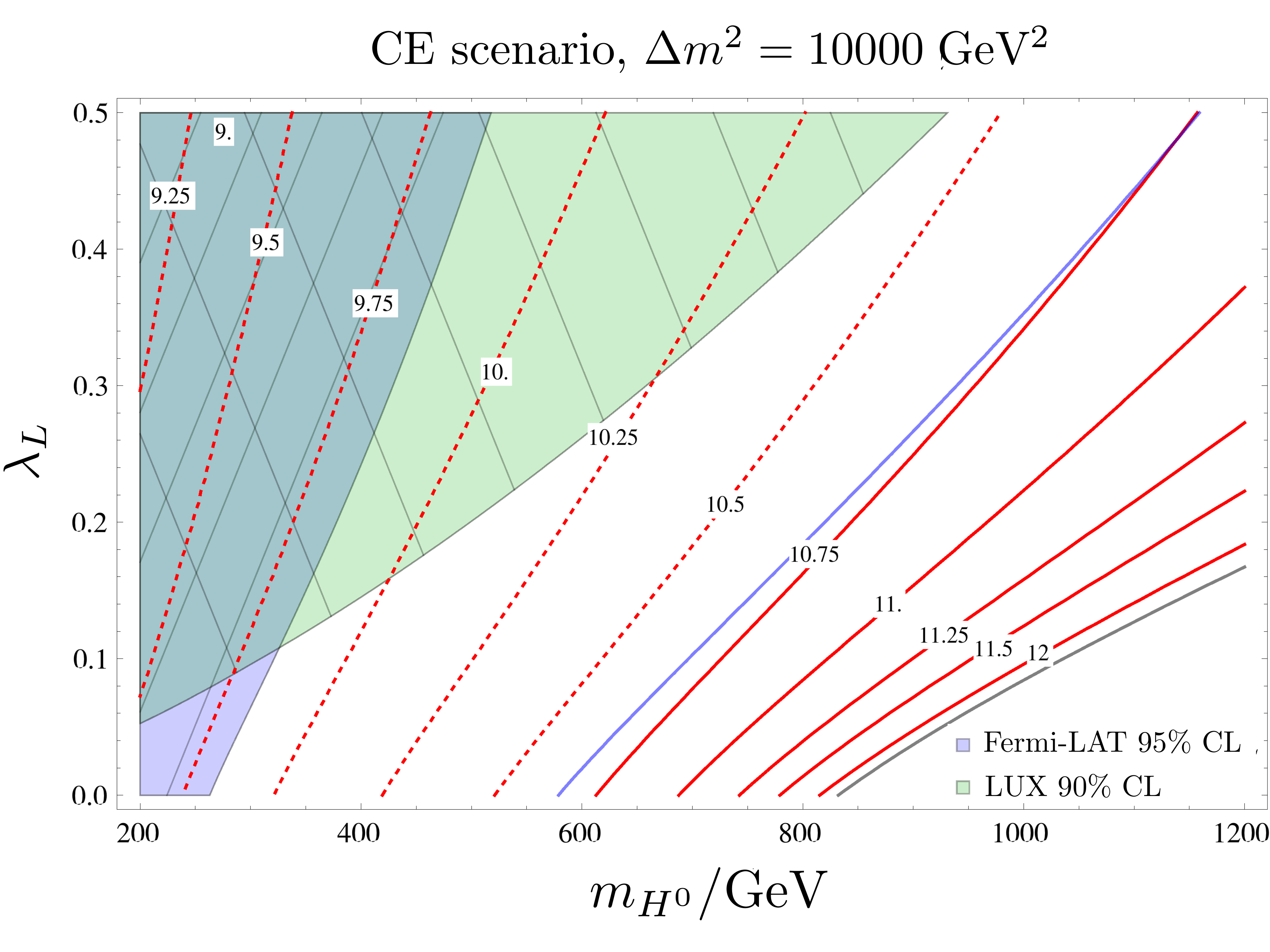}
\caption{Parameter space plots of $\lambda_L$ against $m_{H^0}$ for $\Delta m^2 = 1000 \text{ GeV}^2$ (left) and $10000 \text{ GeV}^2$ (right), in the FT (top) and CE (bottom) scenarios. The green/blue shaded regions are ruled out by LUX/Fermi-LAT. The contour lines correspond to lines of constant $-\log_{10}(q_0)$ values required to produce the observed DM abundance. Of these the red dashed (solid) lines are excluded (allowed) by DGRB and BBN in the FT and CE scenarios respectively, with the blue line representing the upper limit on $q_0$ in each case. Finally, the grey dashed lines represent the limiting case of standard thermal freeze-out.} 
\label{laLMH0Plot}
\end{center}
\end{figure}

We can move away from the benchmark points by allowing the parameter $\la_L$ to vary whilst keeping our two choices of $\Delta m^2$. The plots in Figure~\ref{laLMH0Plot} show the contours of constant $q_0$ (plotted as $-\log_{10}(q_0)$) in the $(m_{H^0},\la_L)$ plane yielding the observed value of the \dm relic density for both choices of $\Delta m^2$ in the FT and CE scenarios. The constraints from DGRB and BBN on the maximum value of $q_0$ have been applied in the FT and CE scenarios respectively. The blue line in each plot indicates the maximum value of $q_0$ allowed by these constraints, with the red dashed lines indicating values excluded by DGRB or BBN with the solid red line indicating allowed values. Also shown in each plot is the freeze-out only case, which is depicted as a grey contour line. 

In addition, the limits from Fermi-LAT/LUX can be applied and the excluded regions of ($m_{H^0}$, $\la_L$) parameter space is indicated by the shaded blue/green areas. Despite these constraints we can see that by introducing the defects into the model we can now generate the correct relic density in regions well beyond the freeze-out only line.  In particular, we note that we can move to lighter dark matter states with possible values down to $\sim 200$ GeV. 

In addition, we note that when we increase the value of $\la_L$ in the small mass splitting case in the CE scenario (bottom left plot of the Figure~\ref{laLMH0Plot}) the origin of the leading limit on $q_0$ changes from BBN to the direct detection experiment LUX. 

Given our limits on $q_0$, one can derive limits on the string tension parameter $G\mu$, (see Appendix \ref{GmuExpressions} for the details of this conversion in the FT and CE cases) for our four benchmark points, these are shown in Table \ref{GmuLimits}. We highlight that the \dm phenomenological limits we can now impose on $G\mu$ (by incorporating \dm freeze-out with defect decays) are stronger than the usual experimental constraints on cosmic strings with large Higgs condensates in the FT scenario. In the CE scenario BBN Higgs condensate constraints are larger than the Fermi-LAT and LUX bounds.

\begin{table}
\begin{center}
\renewcommand*{\arraystretch}{1.2}
\begin{tabular}{|c|c|c|}
\hline
\multirow{2}{*}{$(\la_L,\De m^2/\text{GeV}^2)$} & \multicolumn{2}{|c|}{$G\mu$} \\ \cline{2-3}
 & FT & CE \\ \hline
 $(0,1000)$ & $\lesssim 6\times 10^{-12} P_\text{FT}^{-1}$ & $\lesssim 1\times 10^{-15}P_\text{CE}^{-1}$ \\ \hline
  $(0.1,1000)$ & $\lesssim 3\times 10^{-12} P_\text{FT}^{-1}$ & $\lesssim 1\times 10^{-15}P_\text{CE}^{-1}$ \\ \hline
   $(0,10000)$ & $\lesssim 7\times 10^{-12} P_\text{FT}^{-1}$ & $\lesssim 1\times 10^{-15}P_\text{CE}^{-1}$ \\ \hline
    $(0.1,10000)$ & $\lesssim 7\times 10^{-12} P_\text{FT}^{-1}$ & $\lesssim 1\times 10^{-15}P_\text{CE}^{-1}$ \\ \hline
\end{tabular}
\end{center}
\caption{Limits on $G\mu$ for the chosen benchmark $\la_L$, $\De m^2$ values in the FT and CE scenarios. $P_\text{FT}$ and $P_\text{CE}$ are defined in Appendix \ref{GmuExpressions}.}
\label{GmuLimits}
\end{table}
%%%%%%%%%%%%%%%%%%%%%%%%%%%%%%%%%%%%%%%%%%%%%%%%%%%%%%%%

\section{Conclusion}\label{sec:cons}
%%%%%%%%%%%%%%%%%%%%%%%%%%%%%%%%%%%%%%%%%%%%%%%%%%%%%%%%
In this paper we have studied the effect on the IDM of a new mechanism for the production of \dm from TD decays \cite{Hindmarsh:2013jha}. The defects in question are cosmic strings, produced when an additional $U(1)'$ local symmetry is spontaneously broken in the early universe. Such a $U(1)$ can be a natural extension of the IDM, neatly accommodating the discrete symmetry needed to stabilise the lightest inert CP-even Higgs $H^0$, which plays the role of the \dm in this model. Alternatively, we may expect additional Abelian gauge symmetries from a top down perspective. It is common in theories that are derived from string theory to predict additional light $U(1)$s. These in general must be spontaneously broken and during the resulting phase transition we may expect the formation of cosmic strings. The decay of these TDs can modify the generation of the \dm relic abundance and it is this effect that we have investigated in the context of a test case scenario, the IDM.

The parameters of the IDM are tightly constrained by the requirement to achieve the observed relic density. However, with the new source of \dm states, parameterised by a dimensionless energy injection rate $q_0$, regions of the parameter space which normally under-produce dark matter states can be brought into agreement with the data by an appropriate value of $q_0$. 

Our detailed study of the IDM characterised it with three parameters to allow a comprehensive view; the mass of the dark matter particle $m_{H^0}$, a certain combination of the inert doublet's quartic couplings $\la_L$, and the mass squared splitting between the CP-even the CP-odd inert doublet states $\De m^2$. Additionally we have assumed that the CP-odd and charged inert doublet states were degenerate. 
Within the context of four benchmark points, we took into account bounds on the remaining parameters of the model from direct and indirect detection, from mono-jet searches and the $h\to\ga\ga$ signal strength, and bounds on the energy injection rate from strings in the early universe from big bang nucleosynthesis and the diffuse $\ga$-ray background. It was found that the IDM can accommodate values for the mass of the dark matter particle as low as 220 GeV, more than twice lower than the freeze-out only case.

At the same time, we found tight constraints on the dimensionless string tension parameter $G\mu$ in the IDM model with strings.  If particles are the dominant decay channel of the strings, the limits on $G\mu$ are in the range 3 -- 7 $\times 10^{-12}$, corresponding to an upper bound on the U(1)$'$ symmetry breaking scale of around $10^{13}$ GeV. If gravitational waves are the dominant decay channel, about $G\mu \lesssim10^{-15}$, corresponding to an upper bound on the symmetry-breaking scale of about $10^{12}$ GeV.  The stronger limits in this case are due to the higher length density of string.

Dark matter production by decays of topological defects is relevant for many \dm models that under-predict the value of the relic abundance. One of the immediate implications of such scenarios is that the predicted rates in direct and indirect detection are generically enhanced. This can be a useful in explaining anomalies in experimental data, but it can also place restrictions on the viable parameter space, as we have seen in this work. In any case, it is clear is that spontaneous symmetry breaking of an extra $U(1)$ in the early Universe, can have significant effects in the \dm sector.

\acknowledgments

We acknowledge support from the Science and Technology Facilities Council (grant numbers ST/J000477/1, ST/L000504/1, ST/J000485/1 and, ST/L000512/1). SMW thanks the Oxford Physics Department for hospitality during the completion of this work. J.M.N. is supported by the People Programme (Marie curie Actions) of the European Union Seventh Framework Programme (FP7/2007-2013) under REA grant agreement PIEF-GA-2013-625809.

\appendix
%%%%%%%%%%%%%%%%%%%%%%%%%%%%%%%%%%%%%%%%%%%%%%%%%%%%%%%%

\section{Special features of the Inert Doublet Model}
\label{AppA}

\subsection{Custodial symmetry and the $T$ parameter}
\label{AppA1}

In Section \ref{EWPOsec}, it was noted that the corrections to the electroweak $T$ parameter vanishes in the limits either $m_{A^0} = m_{H^\pm}$ or  $m_{H^0} = m_{H^\pm}$. In this section of the appendix, we show has this can be understood in terms of a custodial symmetry.

As discussed in Section \ref{EWPOsec}, instead of expressing the scalar potential (\ref{Potential}) in terms of $H_{1}, H_{2}$, we can introduce the $2\times2$ matrices ${\bf\Phi}_{1}=(i\sigma _{2}H_{1}^{*}, H_{1}), \, {\bf\Phi}_{2}=(i\sigma _{2}H_{2}^{*}, H_{2})$. The scalar potential for the Inert Doublet Model then reads
\beq\label{Potential_2times2}
\begin{split}
V = & -\frac{\mu_1^2}{2}\,\mathrm{Tr}\left[{\bf\Phi}^{\dagger}_{1} {\bf\Phi}_{1} \right] + \frac{\mu_2^2}{2}\,\mathrm{Tr}\left[{\bf\Phi}^{\dagger}_{2} {\bf\Phi}_{2} \right] + 
\frac{\lambda_1}{4}\,\left(\mathrm{Tr}\left[{\bf\Phi}^{\dagger}_{1} {\bf\Phi}_{1} \right] \right)^2 
+ \frac{\lambda_2}{4}\,\left(\mathrm{Tr}\left[{\bf\Phi}^{\dagger}_{2} {\bf\Phi}_{2} \right] \right)^2  \\ &
+ \frac{\lambda_3}{4}\, \mathrm{Tr}\left[{\bf\Phi}^{\dagger}_{1} {\bf\Phi}_{1} \right] \, \mathrm{Tr}\left[{\bf\Phi}^{\dagger}_{2} {\bf\Phi}_{2} \right] 
+ \frac{\lambda_4 + \lambda_5}{16}\, \left( \mathrm{Tr}\left[{\bf\Phi}^{\dagger}_{1} {\bf\Phi}_{2} \right] + \mathrm{Tr}\left[{\bf\Phi}^{\dagger}_{2} {\bf\Phi}_{1} \right] \right)^2 \\ &
- \frac{\lambda_4 - \lambda_5}{16}\, \left( \mathrm{Tr}\left[{\bf\Phi}^{\dagger}_{1} {\bf\Phi}_{2} \sigma_3\right] - \mathrm{Tr}\left[\sigma_3{\bf\Phi}^{\dagger}_{2} {\bf\Phi}_{1} \right] \right)^2.
\end{split}
\eeq
Both ${\bf\Phi}_{1,2}$ transform as bi-doublets of a global symmetry $SU(2)_L \times SU(2)_R$: ${\bf\Phi}_{i} \to L{\bf\Phi}_{i}R$ (with $L \in SU(2)_L$ and $R \in SU(2)_R$). The potential 
(\ref{Potential_2times2}) is invariant under $SU(2)_L \times SU(2)_R$ in the absence of its last term ($\lambda_4 = \lambda_5$), which would then yield $\Delta m^2_{+} = \Delta m^2_{0}$. 

Moreover, we can define ${\tilde{\bf\Phi}}_{2} = {\bf\Phi}_{2} \sigma_3$, which also transforms as a bi-doublet under $SU(2)_L \times SU(2)_R$. By recasting the potential (\ref{Potential_2times2})
in terms of ${\tilde{\bf\Phi}}_{2}$ and ${\bf\Phi}_{1}$, we see that the $SU(2)_L \times SU(2)_R$ global symmetry is also preserved for $\lambda_4 = -\lambda_5$, which would yield 
$\Delta m^2_{+} = 0$. Then both cases $\lambda_4 = \pm\lambda_5$ preserve a {\it custodial} $SU(2)_L \times SU(2)_R$ symmetry and yield a vanishing $T$ parameter: $\Delta T = 0$.

\subsection{The annihilation cross section $\langle\si_{\text{eff}} v \rangle$ in the pure gauge limit}
\label{AppA2}

In Section \ref{sec:boltzsolve}, it was noted that smaller mass splittings between the Inert Doublet states generates larger relic abundances, despite the greater contribution of co-annihilations. In this section we demonstrate in detail how this can be understood in terms of a unitarity-like cancellation in the $H^0$ annihilation cross-section.

For $m_{H^0} > m_V$, the region of interest to us, the main contributions to $H^0\,H^0$ annihilations are (see {\it e.g.} \cite{Hambye:2009pw}):
\begin{itemize}
\item Contact, $s$-channel (Higgs-mediated) and $t,u$-channels (via $A^{0}/H^{\pm}$) annihilation into massive gauge bosons $H^0\,H^0 \to V\, V$. 
 
\item For $m_{H^0} > m_h$: contact, $s$-channel and $t,u$-channels (via $H^{0}$) annihilation into Higgs bosons $H^0\,H^0 \to h\, h$. 
 
\item For $m_{H^0} > m_t$: $s$-channel (Higgs-mediated) annihilation into top-quark pairs $H^0\,H^0 \to \bar{t}\,t$ (annihilation into other fermions is 
suppressed by the small Yukawa couplings).
  
\end{itemize}

\noindent In the limit $\Delta m^2 \ll m^2_{H^0}$ (where $\De m^2$ is defined in Section \ref{sec:u1ext}), the contributions from co-annihilations with $A^{0}/H^{\pm}$ become important. These are, annihilation into gauge bosons $H^0\,A^0 \to W^{\pm}\, W^{\mp}$, $H^0\,H^{\pm} \to Z\,W^{\pm}$, $H^0\,H^{\pm} \to \gamma\,W^{\pm}$, annihilation into fermion pairs $H^0\,A^0 \to \bar{f}\,f$, $H^0\,H^{\pm} \to \bar{f}\,f'$, and annihilation into Higgs and gauge bosons $H^0\,A^0 \to Z\, h$, $H^0\,H^{\pm} \to W^{\pm}\,h$.

We focus on the dominant \dm annihilation into massive gauge bosons $H^0\,H^0 \to V\, V$. The respective amplitudes read 
\beq\label{amplitudes1}
i \, \mathcal{M}_c = i\,\frac{g^2_V}{2}g^{\mu\nu}\epsilon^*_{\mu}(p_3) \epsilon^*_{\nu}(p_4), \quad \quad i \, \mathcal{M}_s = i\,\frac{\lambda_L v^2\,g^2_V}{s-m^2_h}
g^{\mu\nu}\epsilon^*_{\mu}(p_3) \epsilon^*_{\nu}(p_4),
\eeq 
\beq\label{amplitudes2}
i \, \mathcal{M}_t = i\,\frac{g^2_V\,p^{\mu}_1\,p^{\nu}_2}{t- \Delta m^2 - m^2_{H^0}}\epsilon^*_{\mu}(p_3) \epsilon^*_{\nu}(p_4), \quad \quad i \, 
\mathcal{M}_u = i\,\frac{g^2_V\,p^{\mu}_2\,p^{\nu}_1}{u-\Delta m^2 - m^2_{H^0}}
\epsilon^*_{\mu}(p_3) \epsilon^*_{\nu}(p_4),
\eeq 
with $g_V = g \,(g/\mathrm{cos\,\theta_W})$ for $V = W^{\pm}\, (Z)$. For $\lambda_L = 0$, the amplitude $\mathcal{M}_s$ vanishes, as well as those for the annihilations $H^0\,H^0 \to h\, h$
and $H^0\,H^0 \to \bar{f}\,f$. In the relevant $s$-wave limit, $s = 4\,m^2_{H^0}$, $t = u = m^2_{V} - m^2_{H^0}$, the squared amplitude in this case reads, 
\begin{eqnarray}
\label{amplitudes3}
\left| i \, \mathcal{M}_c + i \, \mathcal{M}_t + i\,  \mathcal{M}_u\right|^2 = g^4_V \left[ \frac{4\,m^4_{H^{0}}}{\left(2\,m^2_{H^{0}}+\Delta m^2-m^2_{W}\right)^2}\left(1 - 2\frac{m^2_{H^{0}}}{m^2_{W}} + \frac{m^4_{H^{0}}}{m^4_{W}} \right) \right. \nonumber \\
\left. - \frac{2\,m^2_{H^{0}}}{2\,m^2_{H^{0}}+\Delta m^2-m^2_{W}}\left(1 - 3 \frac{m^2_{H^{0}}}{m^2_{W}} + 2 \frac{m^4_{H^{0}}}{m^4_{W}} \right) +  
\frac{3}{4} - \frac{m^2_{H^{0}}}{m^2_{W}} + \frac{m^4_{H^{0}}}{m^4_{W}} 
\right] = \nonumber \\
g^4_V \left[\frac{1}{\left(1 + \frac{y}{2} - \frac{x}{2} \right)^2}\left(1 - \frac{2}{x} + \frac{1}{x^2} \right) - \frac{1}{\left(1 + \frac{y}{2} - \frac{x}{2} \right)} \left(1 - \frac{3}{x} + \frac{2}{x^2} \right)+
\frac{3}{4} - \frac{1}{x} + \frac{1}{x^2} \right],\nonumber\\
\end{eqnarray}
with $x \equiv m^2_W/m^2_{H^{0}}$ and $y \equiv \Delta m^2/m^2_{H^{0}}$. As is apparent from (\ref{amplitudes3}), for $\Delta m^2 = 0$ a unitarity cancellation occurs which makes (\ref{amplitudes3}) finite in the limit $x \to 0$
yielding a well-behaved squared amplitude
\beq
\left| \mathcal{M}(H^0\,H^0 \to V\, V) \right|^2_{\De m^2=0} \to  \frac{g^4_V}{2}\,.
\eeq 
For $\Delta m^2 > 0$, the contribution from $t$ and $u$-channels gets suppressed compared to the contact interaction, and the cancellation is only recovered in the limit that $m_{H^0}^2 \gg \De m^2$ ($y \to 0$). In the regime $m_{H^0}^2 \gg \Delta m^2 > m_W^2$, (\ref{amplitudes3}) then reads 
\beq
\label{amplitudes3limit2}
\left| i \, \mathcal{M}_c + i \, \mathcal{M}_t + i\,  \mathcal{M}_u\right|^2 \longrightarrow g^4_V \left[\frac{(\Delta m^2)^2}{4\,m_W^4} 
+ \frac{1}{2} + \mathcal{O}(x,y)\right],
\eeq
which may result in a very large annihilation cross section for $\Delta m^2 \gg m_W^2$ despite the fact that co-annihilations are very suppressed in this case.
Setting $\lambda_L > 0$ reinforces this picture, while for $\lambda_L < 0$ and $\Delta m^2 \gg m^2_{H^0}$ the squared amplitude in the $s$-wave limit $s = 4\,m^2_{H^0}$ reads
\beq\label{amplitudes4}
\left| i \, \mathcal{M}_c + i \, \mathcal{M}_s\right|^2 = \frac{g^4_V}{4}\,\left(2 + \frac{\left(4\,m^2_{H^0} - 2\,m^2_V \right)^2}{4\,m^4_V} \right)\,\frac{\left(4\,m^2_{H^0} - m^2_h - 2\,|\lambda_L|\, v^2\right)^2}{\left(4\,m^2_{H^0} - m^2_h \right)^2} \, ,
\eeq 
and the interplay between the contact and $s$-channel contributions can give rise to a destructive interference, and thus a reduced annihilation cross section \cite{LopezHonorez:2010tb}.

\section{Energy injection rates from cosmic strings}\label{GmuExpressions}
%%%%%%%%%%%%%%%%%%%%%%%%%%%%%%%%%%%%%%%%%%%%%%%%%%%%%%%%%%%%%%%%%%%%%%%%%%%%%%%

In this appendix, we summarise useful formulae for the dark matter energy injection rate from decaying topological defects in the two main cosmic string evolution scenarios. These are the field theory (FT) scenario, in which all energy goes into particles, and the cusp emission (CE) scenario, where particles are emitted only at cusps. More details can be found in Ref. \cite{Hindmarsh:2013jha}.

In the FT scenario, covariant energy conservation dictates that a network of topological defects with energy density $\rho_\text{d}$ will inject energy at a rate per unit volume $Q$ given by  
\beq
{Q} = 3H(w - w_\text{d})\Om_\text{d}{\rho},
\eeq
where $H$ is the Hubble parameter, $\rho$ is the total energy density, $\Om_\text{d} = \rho_\text{d}/\rho$, $w$ is the total average equation of state and $w_\text{d}$ is the defect's average equation of state parameter. For strings with energy per unit length $\mu$, we can define a length scale $\xi = \sqrt{\mu/\rho_\text{d}}$, which can be interpreted as the average distance between strings. We take $\xi \simeq 0.25 d_\text{h}$ in accordance with numerical simulations \cite{Bevis:2006mj,Bevis:2010gj}, where $d_\text{h}$ is the horizon distance. Applying this one finds that the dimensionless energy injection rate into $H^0$ particles, defined by dividing the energy density injection rate by $\rho H$, is
\beq
q_0 = \frac{8\pi G\mu}{\xi^2 H^2}\left(\sum_i \frac{\Ebar_{H^0}}{\Ebar_i} f_i\right)(w - w_\text{d}),
\eeq
where a fraction $f_i$ goes into particles of species $i$ with average energy $\bar E_i$. The decays of each of these particles is assumed to produce one $H^0$.  

Taking $\xi H \simeq 0.25$ and $w_\text{d} \gtrsim -1/3$, one may rewrite this as
\beq \label{q0inFT}
q_0 = \frac{256\pi}{3}G\mu P_\text{FT},
\eeq
where $P_\text{FT}$ is an $\mathcal{O}(1)$ parameter given by
\beq
P_\text{FT} \equiv \left(\sum_i \frac{\Ebar_{H^0}}{\Ebar_i} f_i\right) \left(\frac{1-3w_\text{d}}{2}\right)\left(\frac{0.25}{\xi H}\right)^2.
\eeq
In the CE scenario, the energy density injection rate in the radiation dominated era is given by
\beq
Q_c^{(r)} = \frac{\nu \be_\text{ce}}{\be^{1/2}}\frac{\mu}{t}C\left(\frac{t}{t_\text{ce}}\right),
\eeq
where\footnote{Note that there is an error in \cite{Hindmarsh:2013jha}, where the function $C(\tau)$ is incorrectly given as $3.0\tau^\frac16$ for $\tau \ll 1$.}
\ben
C(\tau) \simeq
\left\{ 
	\ba{cc}
		1.0 \, \tau^\frac16, & \text{for} \; \tau \ll 1,  \cr
		\frac{4}{3} \tau^{-\frac12}, & \text{for} \;  \tau \gg 1.  \cr
	\ea
\right.
\label{e:QcuspNG}
\een
If $t < t_\text{ce}$ cusp emission is more important, as loops have passed a critical size. The parameter $\be_\text{ce}$ contains numerical factors and couplings, $\be = \Ga G\mu$ and $\nu$ is an $\mathcal{O}(1)$ parameter. Evaluating further one finds
\beq
\begin{split}
q_0  = \frac{64}{3}\pi (G\mu\be_\text{ce}^{2/3}\nu) \left(\frac{45 \mpl^2}{16\pi^3}\right)^{1/2} \left(\sum_i \frac{\Ebar_{H^0}}{\Ebar_i} f_i\right) \left(\frac{1}{m_{H^0}\sqrt{g_*}}\right)^{1/6}
 \hspace{-2mm}\simeq (2.01\times 10^4)G\mu P_\text{ce},
\end{split}
\eeq
where again $P_\text{ce}$ is an $\mathcal{O}(1)$ parameter defined as
\beq
P_\text{ce} \equiv \be_\text{ce}^{2/3}\nu \left(\sum_i \frac{\Ebar_{H^0}}{\Ebar_i} f_i\right) \left(\frac{500 \text{ GeV}}{m_{H^0}}\right)^{1/6} \left(\frac{100}{g_*}\right)^{1/12}.
\eeq
The dependence on $m_{H^0}$ is weak enough here to be neglected. 

\bibliographystyle{JHEP}
\bibliography{Bibliography}

\end{document}